\documentstyle[11pt,amssymb,epsfig,psfig]{article}

\makeatletter

\@addtoreset{equation}{section}

\makeatother

\begin{document}

\title{Wightman function and Casimir densities on AdS bulk
with application to the Randall--Sundrum braneworld}

\author{Aram A. Saharian\footnote{%
Email: saharyan@server.physdep.r.am}\\
\textit{Department of Physics, Yerevan State
University, 1 Alex Manoogian Str.,}\\
\textit{375049 Yerevan, Armenia}}
\date{\today}
\maketitle

\begin{abstract}
Positive frequency Wightman function and vacuum expectation value
of the energy-momentum tensor are computed  for a massive scalar
field with general curvature coupling parameter subject to Robin
boundary conditions on two parallel plates located on $(D+1)$-
dimensional AdS background. The general case of different Robin
coefficients on separate plates is considered. The mode summation
method is used with a combination of a variant of the generalized
Abel-Plana formula for the series over zeros of combinations of
cylinder functions. This allows us to extract manifestly the parts
due to the AdS spacetime without boundaries and boundary induced
parts. The asymptotic behavior of the vacuum densities near the
plates and at large distances is investigated. The vacuum forces
acting on the boundaries are presented as a sum of the self-action
and interaction forces. The first one contains well-known surface
divergences and needs further regularization. The interaction
forces between the plates are attractive for Dirichlet scalar. We
show that there is a region in the space of parameters defining
the boundary conditions in which the interaction forces are
repulsive for small distances and attractive for large distances.
An application to the Randall-Sundrum braneworld with arbitrary
mass terms on the branes is discussed.

\end{abstract}

\section{Introduction} \label{sec:introd}

Anti-de Sitter (AdS) spacetime is one of the simplest and most
interesting spacetimes allowed by general relativity. Quantum
field theory in this background has been discussed by several
authors (see, for instance, Refs. \cite{Fron74}--\cite{Gold02}).
Much of early interest to AdS spacetime was motivated by the
questions of principle related to the quantization of fields
propagating on curved backgrounds. The importance of this
theoretical work increased when it was realized that AdS spacetime
emerges as a stable ground state solution in extended supergravity
and Kaluza-Klein models and in string theories.  The appearance of
the AdS/CFT correspondence and braneworld models of
Randall-Sundrum type has revived interest in this subject
considerably. The AdS/CFT correspondence (for a review see
\cite{Ahar00}) represents a realization of the holographic
principle and relates string theories or supergravity in the bulk
of AdS with a conformal field theory living on its boundary. It
has many interesting formal and physical facets and provides a
powerful tool to investigate gauge field theories, in particular
QCD. Recently it has been suggested that the introduction of
compactified extra spatial dimensions may provide a solution to
the hierarchy problem between the gravitational and electroweak
mass scales \cite{Arka98,Rand99a,Rand99b}. The main idea to
resolve the large hierarchy is that the small coupling of four
dimensional gravity is generated by the large physical volume of
extra dimensions. These theories provide a novel setting for
discussing phenomenological and cosmological issues related to
extra dimensions. The model introduced by Randall and Sundrum is
particularly attractive. Their background solution consists of two
parallel flat branes, one with positive tension and another with
negative tension embedded in a five dimensional AdS bulk
\cite{Rand99a}. The fifth coordinate is compactified on $S^1/Z_2$,
and the branes are on the two fixed points. It is assumed that all
matter fields are confined on the branes and only the gravity
propagates freely in the five dimensional bulk. In this model, the
hierarchy problem is solved if the distance between the branes is
about 40 times the AdS radius and we live on the negative tension
brane. More recently, scenarios with additional bulk fields have
been considered \cite{Gher00,Poma00,Davo00,Alte01,Hube01}.

In the scenario presented in \cite{Rand99a} the distance between
the branes is associated with the vacuum expectation value of a
massless scalar field, called the radion. This modulus field has
zero potential and consequently the distance is not determined by
the dynamics of the model. For this scenario to be relevant, it is
necessary to find a mechanism for generating a potential to
stabilize the distance between the branes. Classical stabilization
forces due to the non-trivial background configurations of a
scalar field along an extra dimension were first discussed by
Gell-Mann and Zwiebach \cite{Gell84}. With the revived interest in
extra dimensions and braneworlds, as modified version of this
mechanism, which exploits a classical force due to a bulk scalar
field with different interactions with the branes, received
significant attention \cite{Gold99} (and references therein).
Another possibility for the stabilization mechanism arises due to
the vacuum force generated by the quantum fluctuations about a
constant background of a bulk field. The braneworld corresponds to
a manifold with dynamical boundaries and all fields which
propagate in the bulk will give Casimir-type contributions to the
vacuum energy (for the Casimir effect see Refs.
\cite{Most97}--\cite{Milt02}), and as a result to the vacuum
forces acting on the branes. In dependence of the type of a field
and boundary conditions imposed, these forces can either stabilize
or destabilize the braneworld. In addition, the Casimir energy
gives a contribution to both the brane and bulk cosmological
constants and, hence, has to be taken into account in the
self-consistent formulation of the braneworld dynamics. Motivated
by these, the role of quantum effects in braneworld scenarios has
received some recent attention. For a conformally coupled scalar
this effect was initially studied in Ref. \cite{Fabi00} in the
context of M-theory, and subsequently in Refs.
\cite{Noji00a}--\cite{Knap03} for a background Randall--Sundrum
geometry (for the related heat kernel expansions see Refs.
\cite{Bord99a,Moss00,Gilk01}). The models with dS branes are
considered as well \cite{Eliz03,Noji00b,Nayl02,Moss03}. For a
conformally coupled bulk scalar the cosmological backreaction of
the Casimir energy is investigated in Refs.
\cite{Fabi00,Eliz03,Noji00b,Muko01,Hofm01,Yera03}.

Investigation of local physical characteristics in the Casimir
effect, such as expectation value of the energy-momentum tensor,
is of considerable interest. In addition to describing the
physical structure of the quantum field at a given point, the
energy-momentum tensor acts as the source in the Einstein
equations and therefore plays an important role in modelling a
self-consistent dynamics involving the gravitational field. In
this paper we will study the vacuum expectation value of the
energy-momentum tensor of a scalar field with arbitrary curvature
coupling parameter obeying Robin boundary conditions on two
parallel plates in $(D+1)$-dimensional AdS spacetime.  The general
case is considered when the constants in the Robin boundary
conditions are different for separate plates. Robin type
conditions are an extension of Dirichlet and Neumann boundary
conditions and appear in a variety of situations, including the
considerations of vacuum effects for a confined charged scalar
field in external fields \cite{Ambj83}, spinor and gauge field
theories, quantum gravity and supergravity \cite{Luck91,Espo97}.
Robin conditions can be made conformally invariant, while
purely-Neumann conditions cannot. Thus, Robin-type conditions are
needed when one deals with conformally invariant theories in the
presence of boundaries and wishes to preserve this invariance. It
is interesting to note that the quantum scalar field satisfying
the Robin condition on the boundary of the cavity violates the
Bekenstein's entropy-to-energy bound near certain points in the
space of the parameter defining the boundary condition
\cite{Solo01}. The Robin boundary conditions are an extension of
those imposed on perfectly conducting boundaries and may, in some
geometries, be useful for depicting the finite penetration of the
field into the boundary with the 'skin-depth' parameter related to
the Robin coefficient. Mixed boundary conditions naturally arise
for scalar and fermion bulk fields in the Randall-Sundrum model
\cite{Gher00,Flac01b}. To obtain the expectation values of the
energy-momentum tensor, we first construct the positive frequency
Wightman function. The application of the generalized Abel-Plana
formula to the corresponding mode sum allows us to extract
manifestly the boundary-free AdS part. The expressions for the
boundary induced vacuum expectation values of the energy-momentum
tensor are obtained by applying on the subtracted part a certain
second order differential operator and taking the coincidence
limit. Note that the Wightman function is also important in
consideration of the response of particle detectors at a given
state of motion (see, for instance, \cite{Birrell}).

We have organized the paper as follows. In the next section we
consider the vacuum in the region between two parallel plates. The
corresponding positive frequency Wightman function is evaluated by
using the generalized Abel-Plana summation formula for the series
over zeros of a combination of cylinder functions. This allows us
to present the boundary induced part in terms of integrals with
exponential convergence for the points away the boundaries. In
Section \ref{sec:CD1b} we consider the vacuum expectation value of
the energy-momentum tensor for the case of a single plate. Both
regions on the right and on the left from the plate are
investigated. Various limiting cases are discussed. Section
\ref{sec:2branes} is devoted to the vacuum energy-momentum tensor
for the geometry of two parallel plates. The corresponding vacuum
expectation values are presented in the form of the sum of single
plates and 'interference' parts. The latter is finite everywhere
including the points on the boundaries. The interaction forces
between the plates are discussed as well. In Section
\ref{sec:RanSum} we show that the vacuum expectation value for the
energy-momentum tensor of a bulk scalar in the Randall--Sundrum
braneworld scenario is obtained from our results as a special
case. The last section contains a summary of the work.

\section{Wightman function} \label{sec:WF}

Consider a scalar field $\varphi (x)$ on background of a
$(D+1)$-dimensional plane-symmetric spacetime with the line
element
\begin{equation}\label{metric}
  ds^2=g_{ik}dx^idx^k=e^{-2\sigma (y)}\eta _{\mu \nu }dx^\mu dx^\nu -dy^2,
\end{equation}
and with $\eta _{\mu \nu }={\mathrm{diag}}(1,-1,\ldots ,-1)$ being
the metric for the $D$-dimensional Minkowski spacetime. Here and
below $i,k=0,1,\ldots ,D$, and $\mu ,\nu =0,1,\ldots ,D-1$. The
corresponding field equation has the form
\begin{equation}\label{fieldeq}
  \left( g^{ik}\nabla _i\nabla _k+m^2+\zeta R\right) \varphi (x)=0,
\end{equation}
where the symbol $\nabla _i$ is the operator for the covariant
derivative associated with the metric $g_{ik}$, $R$ is the
corresponding Ricci scalar, and $\zeta $ is the curvature coupling
parameter. For minimally and conformally coupled scalars one has
$\zeta =0$ and $\zeta =\zeta _c=(D-1)/4D$ correspondingly. Note
that by making a coordinate transformation
\begin{equation}\label{zcoord}
  z=\int e^{\sigma (y)}dy,
\end{equation}
metric (\ref{metric}) is written in a conformally-flat form
$ds^2=e^{-2\sigma }\eta _{ik}dx^idx^k$.

Below we will study quantum vacuum effects brought about by the
presence of parallel infinite plane boundaries, located at $y=a$
and $y=b$, $a<b$, with mixed boundary conditions
\begin{equation}\label{boundcond}
    \left( \tilde A_y+\tilde B_y\partial _y\right) \varphi (x)=0,\quad
  y=a,b,
\end{equation}
and constant coefficients $\tilde A_y$, $\tilde B_y$. The presence
of boundaries modifies the spectrum for the zero--point
fluctuations of the scalar field under consideration. This leads
to the modification of the vacuum expectation values (VEVs) of
physical quantities to compared with the case without
boundaries. In particular, vacuum forces arise acting on the
boundaries. This is well known Casimir effect. As a first stage,
in this section we will consider the positive frequency Wightman
function defined as the expectation value
\begin{equation}\label{Wightfunc}
  G^{+}(x,x')=\langle 0|\varphi (x)\varphi (x')|0\rangle ,
\end{equation}
where $|0\rangle $ is the amplitude for the vacuum state. In the
next section we use this function to evaluate the VEVs for the
energy--momentum tensor. Note that the Wightman function also
determines the response of particle detectors in a given state
of motion. By expanding the field operator over eigenfunctions and
using the commutation relations one can see that
\begin{equation}
G^{+}(x,x^{\prime })=\sum_{\alpha }\varphi _{\alpha }(x)\varphi
_{\alpha }^{\ast }(x^{\prime }),  \label{Wightvev}
\end{equation}
where $\alpha $ denotes a set of quantum numbers, and $\left\{
\varphi _{\alpha }(x)\right\} $ is a complete set of solutions to
the field equation (\ref{fieldeq}) satisfying boundary conditions
(\ref{boundcond}).

On the base of the plane symmetry of the problem under
consideration the corresponding eigenfunctions can be presented in
the form
\begin{equation}\label{eigfunc1}
  \varphi _\alpha (x^i)=\phi _{\mathbf{k}}(x^\mu )f_n(y),
\end{equation}
where $\phi _{\mathbf{k}}(x^\mu )$ are the standard Minkowskian
modes on the hyperplane parallel to the plates:
\begin{equation}\label{branefunc1}
  \phi _{\mathbf{k}}(x^\mu )=\frac{e^{-i\eta _{\mu \nu }k^{\mu }x^{\nu }}}{\sqrt{2
  \omega (2\pi )^{D-1}}},\quad k^\mu =(\omega ,{\mathbf{k}}), \quad
  \omega =\sqrt{k^2+m_n^2},\quad k=|{\mathbf{k}}|.
\end{equation}
Here the separation constants $m_n$ are determined by boundary
conditions (\ref{boundcond}) and will be given below. Substituting
eigenfunctions (\ref{eigfunc1}) into the field equation
(\ref{fieldeq}) for the function $f_n(y)$ one obtains the
following equation
\begin{equation}\label{eqforfn}
  -e^{D\sigma }\frac{d}{dy}\left( e^{-D\sigma
  }\frac{df_n}{dy}\right) +\left( m^2+\zeta R\right) f_n=m_n^2e^{2\sigma
  }f_n.
\end{equation}
For the AdS geometry one has $\sigma (y)=k_Dy$, $z=e^{\sigma
(y)}/k_D$, and $R=-D(D+1)k_D^2$, where the AdS curvature radius
is given by $1/k_D$. In this case the solution to equation
(\ref{eqforfn}) for the region $a<y<b$ is
\begin{equation}\label{fny}
  f_n(y)=c_ne^{D\sigma /2}\left[ J_\nu (m_nz)+
  b_\nu Y_\nu (m_nz)\right] ,
\end{equation}
where $J_\nu (x)$, $Y_\nu (x)$ are the Bessel and Neumann functions,
and
\begin{equation}\label{nu}
  \nu =\sqrt{(D/2)^2-D(D+1)\zeta +m^2/k_D^2}.
\end{equation}
Here we will assume values of the curvature coupling parameter for
which $\nu $ is real. For imaginary $\nu $ the ground state
becomes unstable \cite{Brei82}. Note that for a conformally
coupled massless scalar one has $\nu =1/2$ and the cylinder
functions in Eq. (\ref{fny}) are expressed via the elementary
functions. From the boundary condition on $y=a$ one finds
\begin{equation}\label{benu}
  b_\nu =-\frac{\bar J_\nu ^{(a)}(m_nz_a)}{\bar Y_\nu ^{(a)}(m_n
  z_a)}, \quad z_j=e^{\sigma (j)}/k_D,\quad j=a,b\, ,
\end{equation}
where we use the notation
\begin{equation}\label{notbar}
  \bar F^{(j)}(x)=A_jF(x)+B_j xF'(x),\quad
  A_j=\tilde A_j+\tilde B_jk_DD/2, \quad
  B_j=\tilde B_jk_D,\quad j=a,b
\end{equation}
for a given function $F(x)$. Note that for Neumann scalar one has $A_j=B_j D/2$ and
\begin{equation} \label{FNx}
\bar F^{(j)}(x)=F^{{\mathrm{(N)}}}(x)\equiv x^{1-D/2}[x^{D/2}F(x)]',\quad \tilde A_j=0.
\end{equation}
From the boundary condition on the
plane $y=b$ we receive that the eigenvalues $m_n$ have to be
solutions to the equation
\begin{equation}\label{cnu}
  C_\nu ^{ab}(z_b/z_a,m_nz_a)\equiv \bar J_\nu ^{(a)}(m_nz_a)
  \bar Y_\nu ^{(b)}(m_nz_b)-\bar Y_\nu ^{(a)}(m_nz_a)
  \bar J_\nu ^{(b)}(m_nz_b)=0.
\end{equation}
We denote by $z=\gamma _{\nu ,n}$, $n=1,2,\ldots $, the zeros
of the function $C_{\nu }^{ab}(\eta ,z)$ in the right half-plane
of the complex variable $z$, arranged in the ascending order,
$\gamma _{\nu ,n}<\gamma _{\nu ,n+1}$. The eigenvalues for $m_n$
are related to these zeros as
\begin{equation}\label{mntogam}
  m_n=k_D\gamma _{\nu ,n}e^{-\sigma (a)}=\gamma _{\nu ,n}/z_a.
\end{equation}
The coefficient $c_n$ in Eq. (\ref{fny}) is determined from the
orthonormality condition
\begin{equation}\label{ortcond}
  \int _{a}^{b}dye^{(2-D)\sigma }f_n(y)f_{n'}(y)=\delta _{nn'}
\end{equation}
and is equal to
\begin{equation}\label{cn}
  c_n^2=\frac{\pi ^2 u}{2k_Dz_a^2}\bar Y_\nu
  ^{(a)2}(u)T_\nu ^{ab}\left( \eta ,u\right) ,
  \quad u=\gamma _{\nu ,n},\quad \eta =z_b/z_a,
\end{equation}
where we have introduced the notation
\begin{equation}\label{Tnu}
  T_{\nu }^{ab}(\eta ,u)=u\left\{ \frac{\bar J_{\nu }^{(a)2}(u)}{\bar J_{\nu }^{(b)2}
  (\eta u)}\left[ A_b^2+B_b^2(\eta ^2u^2-\nu ^2)\right]-
  A_a^2+B_a^2(u^2-\nu ^2)\right\} ^{-1}.
\end{equation}
Note that, as we consider the quantization in the region between
the branes, $z_a\leq z\leq z_b$, the modes defined by (\ref{fny})
are normalizable for all real values of $\nu $ from Eq.
(\ref{nu}).

Substituting the eigenfunctions (\ref{eigfunc1}) into the mode sum
(\ref{Wightvev}), for the expectation value of the field product
one finds
\begin{equation}
\langle 0|\varphi (x)\varphi (x^{\prime })|0\rangle =
\frac{k_D^{D-1}(z z')^{D/2}}{2^{D+1}\pi ^{D-3}z_a^2}\int d{\mathbf
k}\,e^{i{\mathbf k}({\mathbf x}-{\mathbf x}')}\sum_{n=1}^{\infty
}h_{\nu }(\gamma _{\nu ,n})T_{\nu }^{ab}\left( \eta , \gamma _{\nu
,n}\right) , \label{W11}
\end{equation}
where ${\mathbf x}=(x^1,x^2,\ldots ,x^{D-1})$ represents the
coordinates in $(D-1)$-hyperplane parallel to the plates and
\begin{equation}
h_{\nu }(u)=\frac{ue^{i\sqrt{u^{2}/z_a^{2}+k^{2}}(t^{\prime
}-t)}}{\sqrt{u^{2}/z_a^{2}+k^{2}}}g_{\nu }(u,uz/z_a)g_{\nu
}(u,uz^{\prime }/z_a). \label{hab}
\end{equation}
Here and below we use the notation
\begin{equation}
g_{\nu }(u,v)= J_{\nu }(v)\bar{Y}_{\nu }^{(a)}(u)-\bar{J}_{\nu
}^{(a)}(u)Y_{\nu }(v). \label{genu}
\end{equation}
To sum over $n$ we will use the summation formula derived in
Refs. \cite{Sahreview,Sahsph} by making use of the generalized
Abel-Plana formula \cite{Sahmat,Sahreview}. For a function $h(u)$
analytic in the right half-plane ${\mathrm{Re}}\, u>0$ this
formula has the form
\begin{eqnarray}
\frac{\pi ^{2}}{2}\sum_{n=1}^{\infty }h( \gamma _{\nu ,n})T_{\nu
}(\eta ,\gamma _{\nu ,n}) &=& \int_{0}^{\infty
}\frac{h(x)dx}{\bar{J}_{\nu }^{(a)2}(x)+\bar{Y}_{\nu }^{(a)2}(x)}
- \frac{\pi }{2}{\rm Res}_{u=0}\left[ \frac{h(u)\bar{H}_{\nu
}^{(1b)}(\eta u)}{C_{\nu }^{ab}(\eta ,u)\bar{H}_{\nu
}^{(1a)}(u)}\right] -  \nonumber \\
&&-\frac{\pi }{4}%
\int_{0}^{\infty }\frac{\bar{K}_{\nu }^{(b)}(\eta x)}{\bar{K}_{\nu }^{(a)}(x)%
}\frac{\left[ h(xe^{\pi i/2})+h(xe^{-\pi i/2})\right]
dx}{\bar{K}_{\nu
}^{(a)}(x)\bar{I}_{\nu }^{(b)}(\eta x)-\bar{K}_{\nu }^{(b)}(\eta x)\bar{I}%
_{\nu }^{(a)}(x)}, \label{cor3form}
\end{eqnarray}
where $I_{\nu }(u)$ and $K_{\nu }(u)$ are the Bessel modified
functions. Formula (\ref{cor3form}) is valid for functions $h(u)$
satisfying the conditions
\begin{equation}
|h(u)|<\varepsilon _{1}(x)e^{c_{1}|y|}\quad  |u|\rightarrow \infty
,\quad u=x+iy  , \label{cond31}
\end{equation}
and
\begin{equation}
h(ue^{\pi i})=-h(u)+o(u^{-1}),\quad u\rightarrow 0,
\label{cor3cond1}
\end{equation}
where $c_{1}<2(\eta -1)$, $x^{2\delta _{B_{a}0}-1}\varepsilon
_1(x)\rightarrow 0$ for $x\rightarrow +\infty $. Using the
asymptotic formulae for the Bessel functions for large arguments
when $\nu $ is fixed (see, e.g., \cite{abramowiz}), we can see
that for the function $h_{\nu }(u)$ from Eq. (\ref{hab}) the
condition (\ref{cond31}) is satisfied if $ z+z^{\prime
}+|t-t^{\prime }|<2z_b$. In particular, this is the case in the
coincidence limit $t=t^{\prime }$ for the region under
consideration, $z_a<z,z'<z_b$. As for $|u|<k$ one has $h_{\nu }
(ue^{\pi i})=-h_{\nu }(u) $, the condition (\ref{cor3cond1}) is
also satisfied for the function $h_{\nu}(u)$. Note that $h_{\nu
}(u)\sim u^{1-\delta _{k0}}$ for $u\to 0$ and the residue term on
the right of formula (\ref{cor3form}) vanishes. Applying to the
sum over $n$ in Eq. (\ref{W11}) formula (\ref{cor3form}), one
obtains
\begin{eqnarray}
&&\langle 0|\varphi (x)\varphi (x^{\prime })|0\rangle =
\frac{k_D^{D-1}(z z')^{D/2}}{2^{D}\pi ^{D-1}}\int d{\mathbf
k}\,e^{i{\mathbf k}({\mathbf x}-{\mathbf x}')} \left\{
\frac{1}{z_a^2}\int_{0}^{\infty }\frac{h_{\nu }(u)du}{\bar{J}_{\nu
}^{(a)2}(u)+\bar{Y}
_{\nu }^{(a)2}(u)}-\right.   \label{W13} \\
&&-\left. \frac{2}{\pi }\int_{k }^{\infty }du\, u\frac{\Omega
_{a\nu }(uz_a, uz_b)}{\sqrt{u^{2}-k^2}}G_{\nu }^{(a)}(uz_a,u
z)G_{\nu }^{(a)}(uz_a,uz^{\prime })\cosh \left[
\sqrt{u^{2}-k^{2}}(t-t^{\prime })\right] \right\} , \nonumber
\end{eqnarray}
where we have introduced notations
\begin{eqnarray}
G_{\nu }^{(j)}(u,v)&=&I_{\nu }(v)\bar{K}_{\nu }^{(j)}(u)-\bar{I}%
_{\nu }^{(j)}(u)K_{\nu }(v),\;j=a,b , \label{Geab} \\
\Omega _{a\nu }(u,v) & = & \frac{\bar{K}_{\nu
}^{(b)}(v)/\bar{K}_{\nu }^{(a)}(u)}{\bar{K}_{\nu }^{(a)}(u)\bar{%
I}_{\nu }^{(b)}(v)-\bar{K}_{\nu }^{(b)}(v)\bar{I}_{\nu }^{(a)}(u)}
\label{Omnu}
\end{eqnarray}
(the function with $j=b$ will be used below). Note that we have
assumed values of the coefficients $A_{\alpha }$ and $B_{\alpha }$ for which all zeros
for Eq. (\ref{cnu}) are real and have omitted the residue terms in
the original formula in Refs. \cite{Sahreview,Sahsph}. In the
following we will consider this case only.

To simplify the first term in the figure braces in Eq. (\ref{W13}),
let us use the relation
\begin{eqnarray}
\frac{g_{\nu }(u,uz/z_a)g_{\nu }(u,uz^{\prime }/z_a)}{\bar{J}_{\nu }^{(a)2}(u)+%
\bar{Y}_{\nu }^{(a)2}(u)} & = & J_{\nu }(uz/z_a)J_{\nu }(uz^{\prime }/z_a)-
\nonumber \\
& - & \frac{1}{2} \sum_{\beta =1}^{2}\frac{\bar{J}_{\nu
}^{(a)}(u)}{\bar{H}_{\nu }^{(\beta a)}(u)}H_{\nu }^{(\beta
)}(uz/z_a)H_{\nu }^{(\beta )}(uz^{\prime }/z_a) \label{relab}
\end{eqnarray}
with $H_{\nu }^{(\beta )}(z)$, $\beta =1,2$, being the Hankel
functions. Substituting this into the first integral in the figure
braces of Eq. (\ref{W13}) we rotate the integration contour over
$u$ by the angle $\pi /2$ for $\beta =1$ and by the angle $-\pi
/2$ for $\beta =2$. Under the condition $z+z^{\prime
}-|t-t^{\prime }|>2z_a$, the integrals over the arcs of the circle
with large radius vanish. The integrals over $(0,ikz_a )$ and
$(0,-ikz_a )$ cancel out and after introducing the Bessel modified
functions one obtains
\begin{eqnarray}
\int_{0}^{\infty }\frac{h_{\nu }(u)du}{\bar{J}_{\nu }^{(a)2}(u)+\bar{Y}%
_{\nu }^{(a)2}(u)}& = &  z_a^2\int_{0}^{\infty } du u
\frac{e^{i\sqrt{u^{2}+k^{2}}(t^{\prime
}-t)}}{\sqrt{x^{2}+k^{2}}}J_{\nu }(u z)J_{\nu
}(u z^{\prime })\label{rel1term} \\
&- &\frac{2z_a^2}{\pi }\int_{k}^{\infty }{du\,\
u\frac{\bar{I}_{\nu }^{(a)}(uz_a)}{\bar{K} _{\nu
}^{(a)}(uz_a)}\frac{K_{\nu }(uz)K_{\nu }(uz^{\prime
})}{\sqrt{u^{2}-k^{2} }}\cosh \!}\left[
{\sqrt{u^{2}-k^{2}}(t-t^{\prime })}\right] . \nonumber
\end{eqnarray}
Substituting this into formula (\ref{W13}), the Wightman function
can be presented in the form
\begin{eqnarray}
  \langle 0|\varphi (x)\varphi (x^{\prime })|0\rangle &=&
  \langle 0_S|\varphi (x)\varphi (x^{\prime })|0_S\rangle +
  \langle \varphi (x)\varphi (x^{\prime })\rangle ^{(a)}-
\frac{k_D^{D-1}(z z')^{D/2}}{2^{D-1}\pi ^{D}}\int d{\mathbf
k}\,e^{i{\mathbf k}({\mathbf x}-{\mathbf x}')} \label{W15}
\\
&\times & \int_{k }^{\infty }du u\frac{\Omega _{a\nu }(uz_a,
uz_b)}{\sqrt{u^{2}-k^2}}G_{\nu }^{(a)}(uz_a,uz)G_{\nu
}^{(a)}(uz_a,uz^{\prime })\cosh \left[
\sqrt{u^{2}-k^{2}}(t-t^{\prime })\right] . \nonumber
\end{eqnarray}
Here the term
\begin{equation}\label{WAdS}
  \langle 0_S|\varphi (x)\varphi (x^{\prime })|0_S\rangle =
  \frac{k_D^{D-1}(z z')^{D/2}}{2^{D}\pi ^{D-1}}\int d{\mathbf
k}\,e^{i{\mathbf k}({\mathbf x}-{\mathbf x}')} \int_{0}^{\infty }
du u \frac{e^{i\sqrt{u^{2}+k^{2}}(t^{\prime
}-t)}}{\sqrt{u^{2}+k^{2}}}J_{\nu }(u z)J_{\nu }(u z^{\prime })
\end{equation}
does not depend on the boundary conditions and is the Wightman
function for the AdS space without boundaries.
The second term on the right of Eq. (\ref{W15}),
\begin{eqnarray}
  \langle \varphi (x)\varphi (x^{\prime })\rangle ^{(a)}&=&-
  \frac{k_D^{D-1}(z z')^{D/2}}{2^{D-1}\pi ^{D}}\int d{\mathbf
k}\,e^{i{\mathbf k}({\mathbf x}-{\mathbf x}')} \times \nonumber \\
&\times & \int_{k}^{\infty }{duu\frac{\bar{I}_{\nu
}^{(a)}(uz_a)}{\bar{K} _{\nu }^{(a)}(uz_a)}\frac{K_{\nu }(u
z)K_{\nu }(uz^{\prime })}{\sqrt{u^{2}-k^{2} }}\cosh \!}\left[
{\sqrt{u^{2}-k^{2}}(t-t^{\prime })}\right] , \label{1bounda}
\end{eqnarray}
does not depend on the parameters of the boundary at $z=z_b$ and is induced in the
region $z>z_a$ by a single boundary at $z=z_a$ when the boundary
$z=z_b$ is absent.

Note that expression (\ref{WAdS}) for the boundary-free Wightman
function can also be written in the form
\begin{equation}\label{WAdS1}
  \langle 0_S|\varphi (x)\varphi (x^{\prime })|0_S\rangle =
  k_D^{D-1}(z z')^{D/2} \int_{0}^{\infty }dm \, m G_{M_D}^{+}(x^\mu
  , x'^\mu ; m)J_{\nu }(m z)J_{\nu }(m z^{\prime }),
\end{equation}
where $G_{M_D}^{+}(x^\mu , x'^\mu ;m)$ is the Wightman function
for a scalar field with mass $m$ in the $D$--dimensional Minkowski
spacetime $M_D$. The right hand side in Eq. (\ref{WAdS1}) can be
further simplified by using the expression
\begin{equation}\label{MinkW}
  G_{M_D}^{+}(x^\mu , x'^\mu ;m)=\frac{m^{D/2-1}}{(2\pi )^{D/2}}
  \frac{K_{D/2-1}[m\sqrt{({\mathbf{x}}-{\mathbf{x'}})^2-(t-t'-
  i\varepsilon )^2}]}{[({\mathbf{x}}-{\mathbf{x'}})^2-(t-t'-
  i\varepsilon )^2]^{(D-2)/4}},
\end{equation}
with $\varepsilon >0$. Substituting this into formula
(\ref{WAdS1}) and making use
of the integration formula from \cite{Prudnikov2}, one obtains
\begin{eqnarray}\label{WAdS3}
  \langle 0_S|\varphi (x)\varphi (x^{\prime })|0_S\rangle & = &
  \frac{k_D^{D-1}}{(2\pi )^{(D+1)/2}}e^{-(D-1)\pi i/2}(v^2-1)^{(1-D)/4}
  Q^{(D-1)/2}_{\nu -1/2}(v) \\
  &=& \frac{k_D^{D-1}\Gamma (\nu +D/2)v^{-\nu -D/2}}{2^{D/2+\nu +1}
  \pi ^{D/2}\Gamma (\nu +1)}\,\, {}_2F_1\left( \frac{D+2\nu +2}{4},
  \frac{D+2\nu }{4};\nu +1;\frac{1}{v^2}\right) ,\nonumber
\end{eqnarray}
where $Q^{\alpha }_{\mu }(v)$ is the associated Legendre function
of the second kind, ${}_2F_1(a,b;c;u)$  is the hypergeometric
function (see, for instance, \cite{abramowiz}), and
\begin{equation}\label{ve}
  v=1+\frac{(z-z')^2+({\mathbf{x}}-{\mathbf{x'}})^2-(t-t'-
  i\varepsilon )^2}{2zz'}.
\end{equation}
It can be checked that in the limit $k_D\to 0$ from (\ref{WAdS})
the expression for the $(D+1)$-dimensional Minkowski Wightman
function is obtained. To see this note that in this limit $\nu
\sim m/k_D\to \infty $, $k_Dz\approx 1+k_Dy$ and, hence, as it
follows from (\ref{ve}), one has $v\approx 1+k_D^2w^2/2$, where
$w^2=(y-y^{\prime })^2+({\mathbf{x}}-{\mathbf{x'}})^2-(t-t'-
i\varepsilon )^2$. Using the integral representations for the
functions $Q_{\mu }^{\alpha }(v)$ and $K_{\mu }(v)$ (formulae
8.8.2 and 9.6.23 in Ref. \cite{abramowiz}), it can be seen that
\begin{equation}\label{limQK}
  \lim_{k_D\to 0}\left[ k_D^{2\alpha }e^{-\alpha \pi i}(v^2-1)^{-\alpha /2}
  Q_{\nu -1/2}^{\alpha }(v)\right] =\frac{m^{\alpha }}{w^{\alpha
  }} K_{\alpha }(mw).
\end{equation}
For $\alpha =(D-1)/2$ the expression on the right of this formula
coincides with the $(D+1)$-dimensional Minkowskian Wightman
function up to the coefficient $(2\pi )^{-(D+1)/2}$. Hence, in
combination with the first formula in (\ref{WAdS3}), relation
(\ref{limQK}) proves our statement. Alternatively, we can use
formula (\ref{WAdS}), replacing the Bessel functions by their
asymptotic expressions for large values of the order.

In the coincidence limit $x=x'$ expression (\ref{WAdS3}) is finite
for $D<1$ and for the VEV of the field square one obtains
\begin{equation}\label{phi2AdS}
  \langle 0_S|\varphi ^2(x)|0_S\rangle =\frac{k_D^{D-1}}{(4\pi)^{(D+1)/2}}
  \frac{\Gamma \left( \frac{1-D}{2}\right) \Gamma (D/2+
  \nu )}{\Gamma (1+\nu -D/2)}.
\end{equation}
This quantity is independent on the spacetime point, which is a
direct consequence of the maximal symmetry of the AdS bulk.
Formula (\ref{phi2AdS}) coincides with the result of Refs.
\cite{Burg85,Kame99} (a typo of \cite{Burg85} is corrected in
\cite{Kame99}) obtained from the Feynman propagator in the
coincidence limit (for the zeta function based calculations see
\cite{Camp91,Cald99}). The expression on the right of Eq.
(\ref{phi2AdS}) is analytic in the complex $D$-plane apart from
simple poles coming from the gamma function in the nominator.
Hence, it can be extended throughout the whole complex plane. For
even $D$ this expression is finite and according to the
dimensional regularization procedure \cite{Birrell} can be taken
as the regularized value for the field square. Formula
(\ref{phi2AdS}) can be also obtained directly from (\ref{WAdS})
firstly integrating over ${\mathbf{k}}$ and by making use of the
integration formula
\begin{equation}\label{calI}
{\cal I}_\nu (D)\equiv \int _{0}^{\infty }dx x^{D-1}J_{\nu
}^2(x)=\frac{\Gamma \left( \frac{1-D}{2}\right) \Gamma \left(
D/2+\nu \right)}{2\sqrt{\pi }\Gamma (1-D/2)\Gamma (1+\nu -D/2)}.
\end{equation}

By using the identity
\begin{eqnarray}
&&\frac{\bar{I}_{\nu }^{(a)}(uz_a)}{\bar{K} _{\nu
}^{(a)}(uz_a)}K_{\nu }(u z)K_{\nu }(uz^{\prime })+ \Omega _{a\nu
}(uz_a, uz_b)G_{\nu }^{(a)}(uz_a,u z)G_{\nu
}^{(a)}(uz_a,uz^{\prime })= \nonumber \\
&& \frac{\bar{K}_{\nu }^{(b)}(uz_b)}{\bar{I} _{\nu
}^{(b)}(uz_b)}I_{\nu }(u z)I_{\nu }(uz^{\prime })+ \Omega _{b\nu
}(uz_a, uz_b)G_{\nu }^{(b)}(uz_b,u z)G_{\nu
}^{(b)}(uz_b,uz^{\prime }), \label{ident11}
\end{eqnarray}
with
\begin{equation}\label{Omnub}
\Omega _{b\nu }(u,v) =  \frac{\bar{I}_{\nu
}^{(a)}(u)/\bar{I}_{\nu }^{(b)}(v)}{\bar{K}_{\nu }^{(a)}(u)\bar{%
I}_{\nu }^{(b)}(v)-\bar{K}_{\nu }^{(b)}(v)\bar{I}_{\nu
}^{(a)}(u)},
\end{equation}
it can be seen that the Wightman function in the region $z_a\leq
z\leq z_b$ can also be presented in the equivalent form
\begin{eqnarray}
\langle 0|\varphi (x)\varphi (x^{\prime })|0\rangle &=& \langle
0_S|\varphi (x)\varphi (x^{\prime })|0_S\rangle + \langle \varphi
(x)\varphi (x^{\prime })\rangle ^{(b)}- \frac{k_D^{D-1}(z
z')^{D/2}}{2^{D-1}\pi ^{D}}\int d{\mathbf k}\,e^{i{\mathbf
k}({\mathbf x}-{\mathbf x}')} \label{W17}
\\
&\times & \int_{k }^{\infty }du u\frac{\Omega _{b\nu }(uz_a,
uz_b)}{\sqrt{u^{2}-k^2}}G_{\nu }^{(b)}(uz_b,u z)G_{\nu
}^{(b)}(uz_b,uz^{\prime })\cosh \left[
\sqrt{u^{2}-k^{2}}(t-t^{\prime })\right] , \nonumber
\end{eqnarray}
where
\begin{eqnarray}
\langle \varphi (x)\varphi (x^{\prime })\rangle ^{(b)}&=&-
\frac{k_D^{D-1}(z z')^{D/2}}{2^{D-1}\pi ^{D}}\int d{\mathbf
k}\,e^{i{\mathbf k}({\mathbf x}-{\mathbf x}')} \nonumber \\
&\times & \int_{k}^{\infty }{duu\frac{\bar{K}_{\nu
}^{(b)}(uz_b)}{\bar{I} _{\nu }^{(b)}(uz_b)}\frac{I_{\nu
}(uz)I_{\nu }(uz^{\prime })}{\sqrt{u^{2}-k^{2} }}\cosh \!}\left[
{\sqrt{u^{2}-k^{2}}(t-t^{\prime })}\right] \label{1boundb}
\end{eqnarray}
is the boundary part induced in the region $z<z_b$ by a single
plate at $z=z_b$ when the plate $z=z_a$ is absent. Note that in
the formulae given above the integration over angular part can be
done by using the formula
\begin{equation}\label{angint}
  \int
  d{\mathbf{k}}\, e^{i{\mathbf{k}}({\mathbf{x}}-
  {\mathbf{x'}})}F(k)=(2\pi )^{(D-1)/2}\int _{0}^{\infty }dk\,
  k^{D-2}F(k)\frac{J_{(D-3)/2}(k|{\mathbf{x}}-
  {\mathbf{x'}}|)}{(k|{\mathbf{x}}- {\mathbf{x'}}|)^{(D-3)/2}},
\end{equation}
for a given function $F(k)$. Combining two forms, formulae
(\ref{W15}) and (\ref{W17}), we see that the expressions for the
Wightman function in the region $z_a\leq z \leq z_b $ is symmetric
under the interchange $a \rightleftarrows b $ and $I_{\nu
}\rightleftarrows K_{\nu }$. Note that the expression for the
Wightman function is not symmetric  with respect to the
interchange of the plate indices. The reason for this is that,
though the background AdS spacetime is homogeneous, the boundaries
have nonzero extrinsic curvature tensors and two sides of the
boundaries are not equivalent. In particular, for the geometry of
a single brane the VEVs differ for the regions on the left and on
the right of the brane. Here the situation is similar to that for
the case of a spherical shell on background of the Minkowski
spacetime.

\section{Casimir densities for a single plate} \label{sec:CD1b}

In this section we will consider the VEV of the energy-momentum
tensor for a scalar field in the case of a single plate
located at $z=z_a$. As it has been shown in the previous section
the Wightman function for this geometry is presented in the form
\begin{equation}\label{Wigh1b}
  \langle 0|\varphi (x)\varphi (x')|0\rangle =
  \langle 0_S|\varphi (x)\varphi (x')|0_S\rangle +
  \langle \varphi (x)\varphi (x')\rangle ^{(a)},
\end{equation}
where $|0\rangle $ is the amplitude for the corresponding vacuum
state. The boundary induced part $\langle \varphi (x)\varphi
(x')\rangle ^{(a)}$ is given by formula (\ref{1bounda}) in the
region $z>z_a$ and by formula (\ref{1boundb}) with replacement
$z_b\to z_a$ in the region $z<z_a$. For points away from the plate
this part is finite in the coincidence limit and in the
corresponding formulae for the Wightman function we can directly
put $x=x'$. Introducing a new integration variable
$v=\sqrt{u^{2}-k^{2}}$, transforming to the polar coordinates in
the plane $(v,k)$ and integrating over angular part, the following
formula can be derived
\begin{equation}
\int_{0}^{\infty }dkk^{D-2}\int_{k}^{\infty }du\frac{uf(u)}{\sqrt{u^{2}-k^{2}%
}}=\frac{\sqrt{\pi }\Gamma \left( \frac{D-1}{2}\right) }{2\Gamma (D/2)}%
\int_{0}^{\infty }du u^{D-1}f(u).  \label{rel3}
\end{equation}
By using this formula and Eq. (\ref{1bounda}), the boundary
induced VEV for the field square in the region $z>z_a$ is
presented in the form
\begin{equation}
  \langle \varphi ^2(x)\rangle ^{(a)}=-
  \frac{k_D^{D-1}z^{D}}{2^{D-1}\pi ^{D/2}
  \Gamma (D/2)}\int _{0}^{\infty }du\, u^{D-1}\frac{\bar{I}_{\nu
}^{(a)}(uz_a)}{\bar{K} _{\nu }^{(a)}(uz_a)}K^2_{\nu }(uz), \quad
z>z_a. \label{phi2spl}
\end{equation}
The corresponding formula in the region $z<z_a$ is obtained from
Eq. (\ref{1boundb}) by a similar way and differs from Eq.
(\ref{phi2spl}) by replacements $I_{\nu }\rightleftarrows K_{\nu
}$.

By using the field equation, the expression for the metric energy-momentum
tensor of a scalar field can be presented in the form
\begin{equation}
T_{ik}=\partial _{i}\varphi \partial _{k}\varphi +\left[ \left( \zeta -\frac{%
1}{4}\right) g_{ik}\nabla _{l}\nabla ^{l} -\zeta \nabla _{i}\nabla
_{k}-\zeta R_{ik}\right] \varphi ^{2}. \label{EMT2}
\end{equation}
By virtue of this, for the VEV of the energy-momentum
tensor we have
\begin{equation}
\langle 0|T_{ik}(x)|0\rangle =\lim_{x^{\prime }\rightarrow
x}\partial _{i}\partial _{k}^{\prime }\langle 0|\varphi (x)\varphi
(x^{\prime })|0\rangle +\left[ \left( \zeta -\frac{1}{4}\right)
g_{ik}\nabla _{l}\nabla ^{l} -\zeta \nabla _{i}\nabla _{k}-\zeta
R_{ik}\right] \langle 0|\varphi ^{2}(x)|0\rangle .
\label{vevEMT1pl}
\end{equation}
This corresponds to the point-splitting regularization technique.
VEV (\ref{vevEMT1pl}) can be evaluated by substituting expression
(\ref{Wigh1b}) for the positive frequency Wightman function and
VEV of the field square into Eq. (\ref{vevEMT1pl}). First of all
we will consider the region $z>z_a$. The vacuum energy-momentum
tensor is diagonal and can be presented in the form
\begin{equation}
\langle 0|T_i^k|0\rangle  = \langle 0_S|T_i^k|0_S\rangle + \langle
T_i^k\rangle ^{(a)}, \label{EMT41pl}
\end{equation}
where
\begin{equation}\label{EMTAdS}
  \langle 0_S|T_i^k|0_S\rangle =
  \frac{k_D^{D+1}z^{D}\delta _i^k}{2^{D-1}\pi ^{(D-1)/2}\Gamma
\left( \frac{D-1}{2}\right)}\int _{0}^{\infty } dk
k^{D-2}\int_{0}^{\infty } du u \frac{\tilde f^{(i)}\left[ J_{\nu
}(u z)\right] }{\sqrt{u^{2}+k^{2}}}
\end{equation}
is the VEV for the energy-momentum tensor in the AdS background without boundaries,
and the term
\begin{equation}\label{EMT1bounda}
\langle T_i^k\rangle ^{(a)}=- \frac{k_D^{D+1}z^{D}\delta
_i^k}{2^{D-2}\pi ^{(D+1)/2} \Gamma \left( \frac{D-1}{2}\right)
}\int _{0}^{\infty } d k k^{D-2} \int_{k}^{\infty
}duu\frac{\bar{I}_{\nu }^{(a)}(uz_a)}{\bar{K} _{\nu
}^{(a)}(uz_a)}\frac{\tilde F^{(i)}\left[ K_{\nu }(u z)\right]
}{\sqrt{u^{2}-k^{2}}},
\end{equation}
is induced by a single boundary at $z=z_a$. For a given function
$g(v)$ the functions $\tilde F^{(i)}[g(v)]$ in formula
(\ref{EMT1bounda}) are defined as
\begin{eqnarray}\label{F0}
\tilde F^{(0)}[g(v)] &=& \left( \frac{1}{2}-2\zeta \right) \left[
v^2g'^2(v)+\left( D+\frac{4\zeta }{4\zeta -1}\right) v
g(v)g'(v)+\right.
\\
&& \left. +(v^2+\nu ^2)g^2(v)\right] +(z^2k^2-v^2)g^2(v) \nonumber
\\
\tilde F^{(i)}[g(v)] &=& \tilde F^{(0)}\left[ g(v)\right] +\left( v^2-
\frac{Dz^2k^2}{D-1}\right) g^2(v), \quad i=1,\ldots ,D-1,\label{Fi} \\
\tilde F^{(D)}[g(v)] &=& -\frac{v^2}{2}g'^2(v)+\frac{D}{2}\left(
4\zeta
-1\right) v g(v)g'(v)+ \nonumber \\
&& +\frac{1}{2}\left[ v^2+\nu ^2+2\zeta D(D+1)-D^2/2\right]
g^2(v), \label{FD}
\end{eqnarray}
and the expressions for the functions $\tilde f^{(i)}[g(v)]$ are
obtained from those for $\tilde F^{(i)}[g(v)]$ by the replacement
$v\to iv$. Note that the boundary-induced part (\ref{EMT1bounda})
is finite for the points away the brane and, hence, the
renormalization procedure is needed for the boundary-free part
only. The latter is well-investigated in literature.

Using formula (\ref{rel3}), it can be seen that the contribution
of the second term on the right of Eq. (\ref{Fi}) to the boundary
part of the VEVs vanishes. From Eq. (\ref{EMT1bounda}) now one
obtains
\begin{equation}
\langle T_{i}^{k}\rangle ^{(a)}=-\frac{k_{D}^{D+1}z^{D}\delta
_{i}^{k}}{2^{D-1}\pi ^{D/2}\Gamma (D/2)}\int_{0}^{\infty }du u^{D-1}\frac{%
\bar I_{\nu }^{(a)}(uz_{a})}{\bar K_{\nu
}^{(a)}(uz_{a})}F^{(i)}\left[ K_{\nu }(uz)\right] ,\quad z>z_a,
\label{Tik1plnew}
\end{equation}
where the expressions for the functions $F^{(i)}[g(v)]$ directly
follow from Eqs. (\ref{F0})--(\ref{FD}) after the integration using formula (\ref{rel3}):
\begin{eqnarray}
F^{(i)}[g(v)] &=&\left( \frac{1}{2}-2\zeta \right) \left[
v^2g^{\prime 2}(v)+\left( D+\frac{4\zeta }{4\zeta -1}\right) v
g(v)g^{\prime }(v)+\right.
 \nonumber\\
&&+\left. \left( \nu ^{2}+v^{2}+\frac{2v^{2}}{D(4\zeta -1)}\right)
g^{2}(v)\right] ,\quad i=0,1,\ldots ,D-1, \label{Finew}  \\
F^{(D)}[g(v)] &=&\tilde F^{(D)}[g(v)].  \label{FDnew}
\end{eqnarray}
By a similar way for the VEVs induced by a single brane in the
region $z<z_a$, by making use of expression (\ref{1boundb}) (with
replacement $z_b\to z_a$), one obtains
\begin{equation}
\langle T_{i}^{k}\rangle ^{(a)}=-\frac{k_{D}^{D+1}z^{D}\delta
_{i}^{k}}{2^{D-1}\pi ^{D/2}\Gamma (D/2)}\int_{0}^{\infty }du u^{D-1}\frac{%
\bar K_{\nu }^{(a)}(u z_{a})}{\bar I_{\nu }^{(a)}(u
z_{a})}F^{(i)}\left[ I_{\nu }(u z)\right] ,\quad z<z_a.
\label{Tik1plnewleft}
\end{equation}
Note that VEVs (\ref{Tik1plnew}), (\ref{Tik1plnewleft}) depend
only on the ratio $z/z_a$ which is related to the proper distance
from the plate by equation
\begin{equation}
z/z_a=e^{k_D(y-a)}. \label{propdisz}
\end{equation}
As we see, for the part of the energy-momentum tensor
corresponding to the coordinates in the hyperplane parallel to the
plates one has $\langle T_{\mu \nu }\rangle ^{(a)}\sim \eta _{\mu
\nu }$. Of course, we could expect this result from the problem
symmetry. It can be seen that the VEVs obtained above obey the
continuity equation $T^k_{i;k}=0$ which for the AdS metric takes
the form
\begin{equation}\label{conteq}
  z^{D+1}\frac{\partial }{\partial z}\left( z^{-D}T_D^D\right) +
  D T_0^0=0.
\end{equation}

In the AdS part without boundary the integration over $k$ can be
done using the formula
\begin{equation}
\int_{0}^{\infty }\frac{k^{D-2}dk}{\sqrt{u^{2}+k^{2}}}=\frac{u^{D-2}}{2\sqrt{%
\pi }}\Gamma \left( \frac{D-1}{2}\right) \Gamma \left(
1-\frac{D}{2}\right) . \label{rel5}
\end{equation}
As for the boundary terms, the contribution corresponding to the
second term on the right of Eq. (\ref{Fi}) vanishes and one
obtains
\begin{equation}
\langle 0_{S}|T_{i}^{k}|0_{S}\rangle =\frac{k_{D}^{D+1}z^{D}%
\delta _{i}^{k}}{2^{D}\pi ^{D/2}}\Gamma \left(
1-\frac{D}{2}\right) \int_{0}^{\infty }duu^{D-1}f^{(i)}\left[
J_{\nu }(uz)\right] , \label{AdSnew}
\end{equation}
where the expressions for the functions $f^{(i)}\left[ g(v)\right]
$ are obtained from those for $F^{(i)}\left[ g(v)\right] $, Eqs.
(\ref{Finew}), (\ref{FDnew}), by the replacement $v\to iv$. Now from Eq.
(\ref{Finew}) we have the following relation between the vacuum
energy density and pressures in the parallel directions (no
summation over~$i$)
\begin{equation}\label{enpreseq}
\langle 0|T^{0}_{0}|0\rangle =\langle 0|T^{i}_{i}|0\rangle ,\quad
i=1,\ldots , D-1.
\end{equation}
To evaluate the $u$-integral in (\ref{AdSnew}) we use formula
(\ref{calI}) and the relations
\begin{equation}\label{calIrel}
{\cal I}_\nu (D+2)=\left( \nu ^2-\frac{D^2}{4}\right) \frac{D{\cal
I}_\nu (D)}{D+1}, \quad {\cal I}_{\nu -1} (D+2)=\left( \nu
-1-\frac{D}{2}\right) \left( \nu -\frac{D}{2}\right) \frac{D{\cal
I}_\nu (D)}{D+1},
\end{equation}
\begin{equation}\label{calIrel2}
\int _{0}^{\infty }dx x^{D}J_{\nu }(x)J_{\nu -1}(x)=\left( \nu
-\frac{D}{2}\right) {\cal I}_\nu (D).
\end{equation}
This yields
\begin{equation}\label{fiintads}
\int _{0}^{\infty }dx x^{D-1}f^{(i)}[J_{\nu
}(x)]=\frac{m^2}{k_D^2}\frac{{\cal I}_\nu (D)}{D+1},\quad
i=0,1,\ldots ,D.
\end{equation}
As a result for the boundary-free AdS VEVs one receives
\begin{equation}\label{adsvevpart}
\langle 0_{S}|T_{i}^{k}|0_{S}\rangle =\delta _{i}^{k}
\frac{k_D^{D-1}m^2}{(4\pi )^{(D+1)/2}}\frac{\Gamma \left(
\frac{1-D}{2}\right) \Gamma \left( D/2+\nu \right)}{(D+1)\Gamma
(1+\nu -D/2)}.
\end{equation}
Hence, the AdS VEVs are proportional to the corresponding metric
tensor. Again, we could expect this result due to the maximally
symmetry of the AdS background. Formula (\ref{adsvevpart}) can be
obtained also directly from (\ref{phi2AdS}), by using the standard
relation between the field square and the trace of the
energy-momentum tensor.

For a conformally coupled massless scalar $\nu =1/2$, and by
making use of the expressions for the modified Bessel functions,
it can be seen that $\langle T_{i}^{k}\rangle ^{(a)}=0$ in the
region $z>z_a$ and
\begin{equation} \label{conf1pl}
\langle T_{D}^{D}\rangle ^{(a)}=-D\langle T_{0}^{0}
\rangle ^{(a)}=- \frac{(k_Dz/z_a)^{D+1}}{(4\pi )^{D/2}
\Gamma (D/2)} \int _0^{\infty } \frac{t^D\, dt}{\frac{B _a (t-1)+
2A_a}{B _a (t+1)-2A_a}e^t+1}
\end{equation}
in the region $z<z_a$. Note that the corresponding energy-momentum
tensor for a single Robin plate in the Minkowski bulk vanishes
\cite{Rome02} and the result for the region $z>z_a$ is obtained by
a simple conformal transformation from that for the Minkowski
case.  In the region $z<z_a$ this procedure does not work as in
the AdS problem one has $0<z<z_a$ instead of $-\infty <z<z_a$ in
the Minkowski problem and, hence, the part of AdS under
consideration is not a conformal image of the corresponding
manifold in the Minkowski spacetime (for a general discussion of
this question in conformally related problems see Ref.
\cite{Cand}).

The boundary induced VEVs given by equations (\ref{Tik1plnew}) and
(\ref{Tik1plnewleft}), in general, can not be further simplified
and need numerical calculations. Relatively simple analytic
formulae for the brane-induced parts can be obtained in limiting
cases. First of all, as a partial check, in the limit $k_D\to 0$
the corresponding formulae for a single plate on the Minkowski
bulk are obtained (see Ref. \cite{Rome02} for the massless case
and Ref. \cite{Mate} for the massive case). This can be seen
noting that in this limit $\nu \sim m/k_D$ is large and by
introducing the new integration variable in accordance with $u=\nu
y$, we can replace the Bessel modified functions by their uniform
asymptotic expansions for large values of the order. The Minkowski
result is obtained in the leading order.

In the limit $ z\to z_{a}$ for a fixed $k_D$ expressions
(\ref{Tik1plnew}) and (\ref{Tik1plnewleft}) diverge. In accordance
with (\ref{propdisz}), this corresponds to small proper distances
from the brane. The surface divergences in the VEVs of the
energy-momentum tensor are well-known in quantum field theory with
boundaries and are investigated for various types of boundary
geometries and boundary conditions (see, for instance, Refs.
\cite{Deut79,Kenn80}). Near the brane the main contribution into
the integral over $u$ in Eqs. (\ref{Tik1plnew}),
(\ref{Tik1plnewleft}) comes from the large values of $u$ and we
can replace the Bessel modified functions by their asymptotic
expressions for large values of the argument when the order is
fixed (see, for instance, \cite{abramowiz}). To the leading order
this yields
\begin{eqnarray}
\langle T_{0}^{0}\rangle ^{(a)} &\sim &\Gamma
\left( \frac{D+1}{2}\right) \frac{Dk_{D}^{D+1}(\zeta -\zeta
_{c})\kappa (B_a)}{2^{D}\pi
^{(D+1)/2}|1-z_{a}/z|^{D+1}},
\label{Tik1near}
\\
\langle T_{D}^{D}\rangle ^{(a)} &\sim &\langle T_{0}^{0}\rangle
^{(a)}\left( 1-\frac{z_{a}}{z}\right) \label{TDD1near},
\end{eqnarray}
where we use the notation
\begin{equation} \label{kappa}
\kappa (x)=\left\{ \begin{array}{ll}
1 & {\mathrm{if}} \quad x=0 \, ,\\
-1 & {\mathrm{if}} \quad x\neq 0 \, .
\end{array} \right.
\end{equation}
Note that the leading terms for the components with $i=0,1,\ldots
,D-1$ are symmetric with respect to the plate, and the $_{D}^{D}$
-- component has opposite signs for the different sides of the
plate. Near the brane the vacuum energy densities have opposite
signs for the cases of Dirichlet ($B_a=0$) and non-Dirichlet
($B_a\neq 0$) boundary conditions. Recall that for a conformally
coupled massless scalar the vacuum energy-momentum tensor vanishes
in the region $z>z_a$ and is given by expression (\ref{conf1pl})
in the region $z<z_a$. The latter is finite everywhere including
the points on the plate.

For the points with the proper distances from the plate much
larger compared with the AdS curvature radius one has $z\gg
z_{a}$. This limit is important from the point of view of the
application to the Randall-Sundrum braneworld. Introducing in Eq.
(\ref{Tik1plnew}) a new integration variable $ y=zu$, by making
use of formulae for the Bessel modified functions for small values
of the argument, and assuming $A_{a}-B_{a}\nu \neq 0$, to the
leading order we receive
\begin{equation}
\langle T_{i}^{k}\rangle ^{(a)}\sim -\frac{k_{D}^{D+1}\delta
_{i}^{k}2^{2-D}\pi ^{-D/2}
}{\Gamma (D/2)\Gamma (\nu )\Gamma (\nu +1)}\frac{%
A_{a}+B_{a}\nu }{A_{a}-B_{a}\nu }\left( \frac{z_{a}}{2z}\right)
^{2\nu }\int_{0}^{\infty }dyy^{D+2\nu -1}F^{(i)}\left[ K_{\nu
}(y)\right] . \label{Tik1far}
\end{equation}
The integral in this formula can be evaluated on the base of
formulae
\begin{eqnarray}
\int_{0}^{\infty }dxx^{\alpha -1}K_{\nu }^{2}(x) &=&\frac{2^{\alpha -3}}{%
\Gamma (\alpha )}\Gamma \left( \frac{\alpha }{2}+\nu \right)
\Gamma \left(
\frac{\alpha }{2}-\nu \right) \Gamma ^{2}\left( \frac{\alpha }{2}\right) ,
\label{intK1}\\
\int_{0}^{\infty }dxx^{\alpha -1}K_{\nu }^{^{\prime }2}(x) &=&\left[ \frac{%
\alpha ^{2}}{2}-\nu ^{2}-\frac{\alpha }{\alpha +1}\left( \frac{\alpha ^{2}}{4%
}-\nu ^{2}\right) \right] \int_{0}^{\infty }dxx^{\alpha -1}K_{\nu
}^{2}(x) \label{intK2}.
\end{eqnarray}
This yields
\begin{eqnarray}
\langle T_{0}^{0}\rangle ^{(a)} &\sim &\frac{k_{D}^{D+1}
}{(4\pi )^{(D-1)/2}}\frac{A_{a}+B_{a}\nu }{A_{a}-B_{a}\nu
}\left(
\frac{z_{a}}{2z}\right) ^{2\nu }\left( 4\zeta -\frac{D+2\nu }{D+2\nu +1}%
\right)  \nonumber \\
&&\times \, \frac{(2\nu -1)\Gamma (D/2+\nu +1)\Gamma (D/2+2\nu
)}{\Gamma (\nu
+1)\Gamma (D/2+1/2+\nu )}, \label{Tik1far1}\\
\langle T_{D}^{D}\rangle ^{(a)} &\sim &\frac{D}{D+2\nu }\langle
T_{0}^{0}\rangle ^{(a)},\quad z\gg z_{a}, \label{TDD1far1}
\end{eqnarray}
and the boundary-induced VEVs are exponentially suppressed by the
factor $\exp [2\nu k_D(a-y)]$. In the limit under consideration
the ratio of the energy density and the perpendicular pressure is
a negative constant. We see that in the case $\zeta \leq \zeta
_c$, at large distances from the plate the energy density is
positive for $|B_a/A_a|>1/\nu $ and is negative for
$|B_a/A_a|<1/\nu $. Combining this with the asymptotic behavior
near the plate, Eq. (\ref{Tik1near}), we conclude that for
$B_a\neq 0$ and $|B_a/A_a|<1/\nu $ the energy density is positive
near the plate and is negative at large distances approaching to
zero and, hence, for some value of $z$ it has a minimum with the
negative energy density.

Now we turn to the limit $z\ll z_{a}$ for a fixed $k_D$. This
corresponds to the points near the AdS boundary presented by the
hyperplane $z=0$ with the proper distances from the brane much
larger compared with the AdS curvature radius. Introducing in Eq.
(\ref{Tik1plnewleft}) a new integration variable $y=uz_{a}$ and by
making use of the formulae for the Bessel modified functions for
small values of the argument, to the leading order one receives
\begin{eqnarray}
\langle T_{0}^{0}\rangle ^{(a)} &\sim &-\frac{k_{D}^{D+1}}{\pi ^{D/2}
\Gamma (D/2)}\left( \frac{z}{2z_{a}}\right) ^{D+2\nu }%
\frac{D+2\nu -4\zeta (D+2\nu +1)}{\Gamma (\nu )\Gamma (\nu +1)}  \nonumber \\
&&\times \int_{0}^{\infty }dxx^{D+2\nu -1}
\frac{\bar K_{\nu }^{(a)}(x)}{\bar I_{\nu }^{(a)}(x)}%
, \label{Tik1near0}\\
\langle T_{D}^{D}\rangle ^{(a)} &\sim &-\frac{D}{2\nu }\langle
T_{0}^{0}\rangle ^{(a)},\quad z\ll z_{a} \label{TDD1near0} .
\end{eqnarray}
In this limit the exponential suppression takes place by the
factor $\exp [(D+2\nu )k_D(y-a)]$. The ratio of the energy density
and the perpendicular pressure is a positive constant. Due to the
well-known properties of the Bessel modified functions, the
integral in Eq. (\ref{Tik1near0}) is positive for small values of
the ratio $B_a/A_a$ and is negative for large values and, hence,
as in the previous case there is a possibility for the change of
the sign for the energy density as a function on the distance from
the boundary. Now by taking into account relation
(\ref{propdisz}), from Eqs. (\ref{Tik1far1}), (\ref{Tik1near0}) we
conclude that for large proper distances from the plate to
compared with the AdS curvature radius the boundary induced parts
are exponentially suppressed.

In the large mass limit, $m\gg k_D$, from (\ref{nu}) one has $\nu
\approx m/k_D\gg 1 $. To find the leading terms of the
corresponding asymptotic expansions for the vacuum energy-momentum
tensor components we replace the integration variable in Eqs.
(\ref{Tik1plnew}) and (\ref{Tik1plnewleft}), $x=\nu t$, and use
the uniform asymptotic expansions for the Bessel modified
functions for large values of the order. The main contribution
to the integrals over $t$ comes from small values of $t$. For a
fixed $z$ to the leading order one receives
\begin{eqnarray}\label{Tiklargenu}
  \langle T_0^0\rangle ^{(a)} &\sim & (4\zeta -1)\kappa (B_a)
  \frac{k_D^{D+1}}{\pi ^{D/2}}\left( \frac{m}{2k_D}
  \right) ^{D/2+1}\frac{e^{-2|\ln
  (z_a/z)|m/k_D}}{|1-z_a^2/z^2|^{D/2}},  \\
\langle T_D^D\rangle ^{(a)} &\sim & \frac{Dk_D}{2m}\langle
T_0^0\rangle ^{(a)},\quad m\gg k_D, \label{TDDlargenu}
\end{eqnarray}
and we have an exponential suppression of the vacuum expectation
values for large values of the mass.

For large values of the parameter $k_D$ and fixed proper distances
from the plate, $k_D|y-a|\gg 1$, from relation (\ref{propdisz})
one has $z_a\ll z$ or $z_a\gg z$. Hence, in this limit, which
corresponds to strong gravitational fields, we have an asymptotic
behavior described by Eqs. (\ref{Tik1far1}), (\ref{TDD1far1}) in
the region $z>z_a$ and by Eqs. (\ref{Tik1near0}),
(\ref{TDD1near0}) in the region $z<z_a$ with the exponential
suppression of the boundary induced parts by the factor $\exp
[2\nu k_D(a-y)]$ in the first case and by the factor $\exp [(2\nu
+D)k_D(y-a)]$ in the second case.

\begin{figure}[tbph]
\begin{center}
\begin{tabular}{cc}
\epsfig{figure=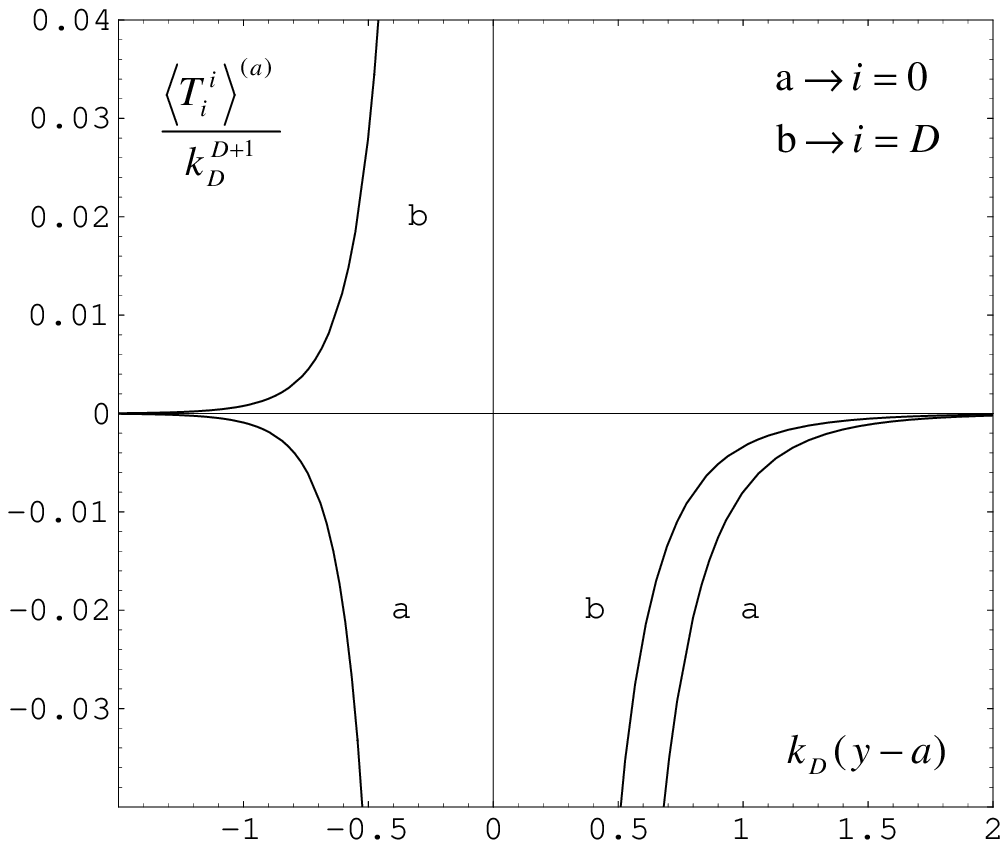,width=6.5cm,height=6cm}& \quad
\epsfig{figure=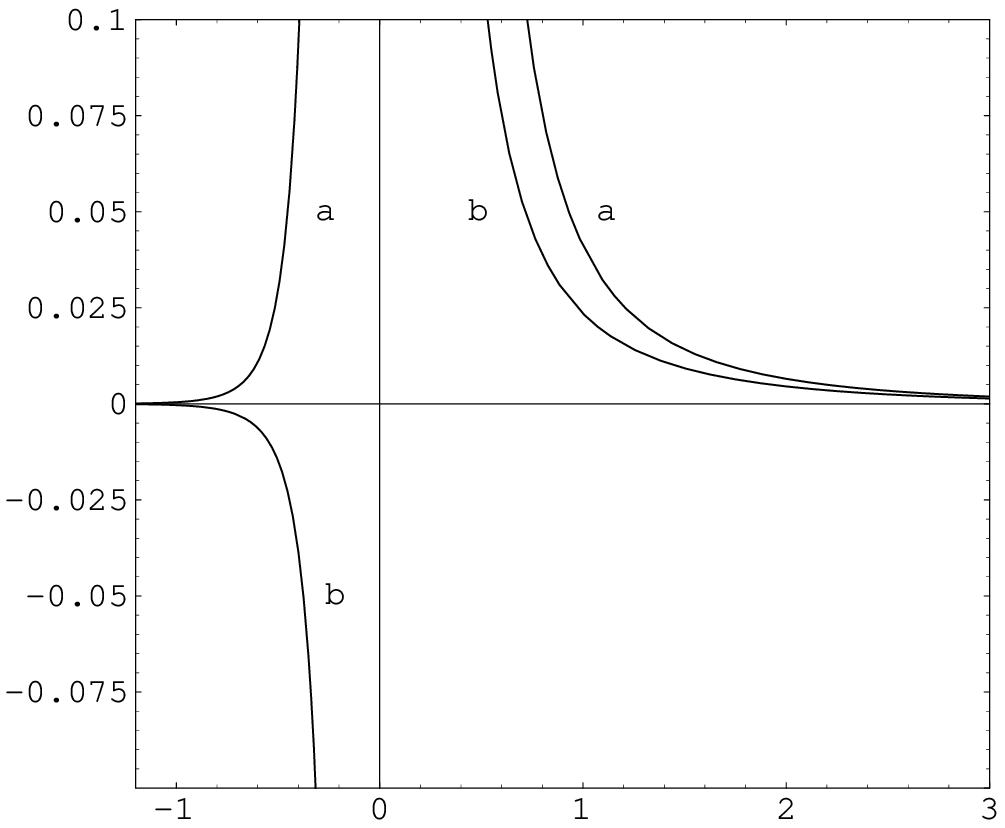,width=6.5cm,height=6cm}
\end{tabular}
\end{center}
\caption{Boundary induced vacuum densities $\langle T^0_0\rangle
^{(a)}/k_{D}^{D+1}$ and $\langle T^D_D\rangle ^{(a)}/k_{D}^{D+1}$
as functions on $k_D(y-a)$ for a minimally coupled massless scalar
in $D=3$. The curves (a) correspond to the energy density and the
curves (b) correspond to the ${}^{D}_{D}$ -- component of the
vacuum stress. The left panel is for Dirichlet scalar and the
right one is for Neumann scalar.} \label{figsdn}
\end{figure}

\begin{figure}[tbph]
\begin{center}
\epsfig{figure=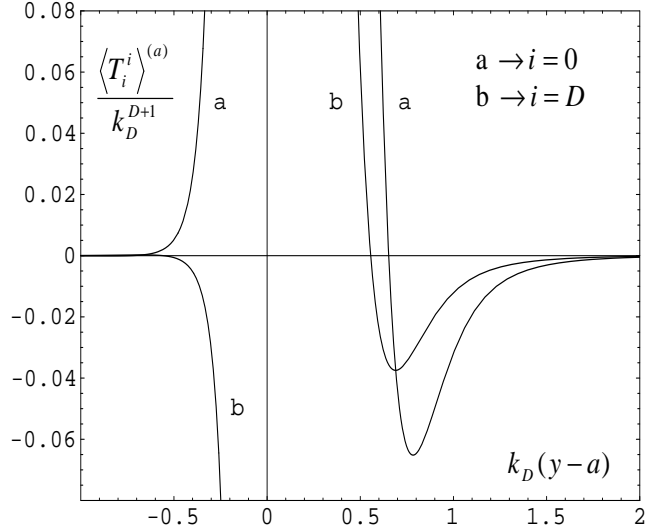,width=8.5cm,height=7cm}
\end{center}
\caption{The same as in figure \ref{figsdn} for Robin scalar with
$B_a/A_a=1/4$.} \label{figsro}
\end{figure}

In figures \ref{figsdn} and \ref{figsro} we have plotted the
vacuum energy density [curves (a)] and ${}^{D}_{D}$ -- stress
[curves (b)] induced by a single plate, as functions on $k_D(y-a)$
for a minimally coupled massless scalar field in $D=3$. In figure
\ref{figsdn} the cases of Dirichlet (left panel) and Neumann
(right panel) boundary conditions are presented. The energy
density is negative everywhere for Dirichlet scalar and is
positive for Neumann scalar. It can be seen that for both
Dirichlet and Neumann cases the boundary-induced energy-momentum
tensors violate the strong energy condition. The weak energy
condition is violated by Dirichlet scalar and is satisfied by
Neumann scalar. The energy conditions play a key role in the
singularity theorems of general relativity and these properties
may have important consequences in the consideration of the
gravitational back reaction of vacuum quantum effects. Figure
\ref{figsro} corresponds to the Robin boundary condition with
$B_a/A_a=1/4$ and illustrates the possibility for the change of
the sign for the energy density as a function on distance from the
plate: the energy density is positive near the plate and is
negative for large distances from the plate. Note that the ratio
$\langle T_D^D\rangle ^{(a)}/\langle T_0^0\rangle ^{(a)}$ is a
coordinate dependent function. In accordance with the asymptotic
estimates given above, this ratio tends to zero for the points
near the brane and to the constant values $D/(D+2\nu )$ and
$-D/(2\nu )$ in the limits $z\gg z_a$ and $z\ll z_a$,
respectively.

\bigskip


\section{Two-plates geometry} \label{sec:2branes}

\subsection{Vacuum densities}

In this section we will investigate the VEVs for the
energy-momentum tensor in the region between two parallel plates.
Substituting the corresponding Wightman function from Eq.
(\ref{W15}) into the mode sum formula (\ref{vevEMT1pl}), we obtain
\begin{eqnarray}
\langle 0|T_i^k|0\rangle & = & \langle 0_S|T_i^k|0_S\rangle +
\langle T_i^k\rangle ^{(a)}- \frac{k_D^{D+1}z^{D}\delta
_i^k}{2^{D-2}\pi ^{(D+1)/2}\Gamma \left( \frac{D-1}{2}\right)}\int
_0^\infty d k k^{D-2}
\label{EMT4} \\
& \times & \int_{k }^{\infty }du u\frac{\Omega _{a\nu }(uz_a,
uz_b)}{\sqrt{u^{2}-k^2}}\tilde F^{(i)}\left[G_{\nu }^{(a)}(uz_a,u
z)\right] , \quad z_a<z<z_b \nonumber
\end{eqnarray}
where for a given function $g(v)$ the functions $\tilde F^{(i)}[g(v)]$
are defined in accordance with Eqs. (\ref{F0})--(\ref{FD}). By
using formula (\ref{rel3}), we see that the contribution of the
second term on the right of Eq. (\ref{Fi}) vanishes and from Eq.
(\ref{EMT4}) one obtains
\begin{equation}
\langle 0|T_{i}^{k}|0\rangle =\langle 0_{S}|T_{i}^{k}|0_{S}\rangle
+\langle T_{i}^{k}\rangle ^{(a)}-\frac{k_{D}^{D+1}z^{D}\delta
_{i}^{k}}{2^{D-1}\pi ^{\frac{D}{2}}\Gamma \left(
\frac{D}{2}\right) } \int_{0}^{\infty }du u^{D-1}\Omega _{a\nu }(u
z_{a},u z_{b})F^{(i)}\left[ G_{\nu }^{(a)}(u z_{a},u z)\right] ,
\label{Tik1int}
\end{equation}
with the functions $F^{(i)}[g(v)]$ from Eqs.
(\ref{Finew}),(\ref{FDnew}).

By making use of the Wightman function in form (\ref{W17}), we
obtain an alternative representation of the VEVs:
\begin{equation}
\langle 0|T_{i}^{k}|0\rangle =\langle 0_{S}|T_{i}^{k}|0_{S}\rangle
+\langle T_{i}^{k}\rangle ^{(b)}-\frac{k_{D}^{D+1}z^{D}\delta
_{i}^{k}}{2^{D-1}\pi ^{\frac{D}{2}}\Gamma \left(
\frac{D}{2}\right) } \int_{0}^{\infty }du u^{D-1}\Omega _{b\nu }(u
z_{a},u z_{b})F^{(i)}\left[ G_{\nu }^{(b)}(u z_{b},u z)\right] .
\label{Tik1intb}
\end{equation}
This form is obtained from (\ref{Tik1int}) by the replacements
$a\rightleftarrows b$, $I_{\nu }\rightleftarrows K_{\nu }$. The
vacuum energy-momentum tensor in the region between the branes is
not symmetric under the interchange of indices of the branes. As
it has been mentioned above, the reason for this is that, though
the background spacetime is homogeneous, due to the non-zero
extrinsic curvature tensors for the branes, the regions on the
left and on the right of the brane are not equivalent. By the same
way as for the case of a single plate, it can be seen that in the
limit $k_D\to 0$ from formulae (\ref{Tik1int}) and
(\ref{Tik1intb}) the corresponding results for two plates geometry
in the Minkowski bulk (see Refs. \cite{Rome02,Mate}) are obtained.

On the base of formula (\ref{Tik1int}), the VEVs in the region
$z_{a}<z<z_{b}$ can be written in the form
\begin{equation}\label{Tik2plint}
\langle 0|T_{i}^{k}|0\rangle =\langle 0_S|T_{i}^{k}|0_S\rangle +
\langle T_{i}^{k}\rangle ^{(a)}+\langle T_{i}^{k}\rangle ^{(b)}+
\Delta \langle T_{i}^{k}\rangle ,
\end{equation}
with the 'interference' term
\begin{eqnarray}
\Delta \langle T_{i}^{k}\rangle &=& -\frac{k_D^{D+1}z^D\delta
_i^k}{2^{D-1}\pi ^{D/2}\Gamma (D/2)}\int _{0}^{\infty }du\,
u^{D-1} \times \nonumber \\
&& \times \left\{ \Omega _{a\nu }(uz_a,uz_b)F^{(i)}[G_{\nu
}^{(a)}(uz_a,u z)]-\frac{\bar K_\nu ^{(b)}(uz_b)}{\bar I_\nu
^{(b)}(uz_b)}F^{(i)}[I_{\nu }(u z)]\right\} .\label{intTik1}
\end{eqnarray}
Another form we can obtain by using formulae (\ref{Tik1plnew}) and
(\ref{Tik1intb}). Note that the surface divergences are contained
in the second and third terms on the right of Eq.
(\ref{Tik2plint}) and expression (\ref{intTik1}) is finite for all
values $z_a\leq z\leq z_b$ (for large values $u$ the subintegrand
behaves as $u^{D}e^{2u(z_{a}-z_{b})}$). For the proper distances
between the plates much less than the AdS curvature radius,
$k_D(b-a)\ll 1$, one has $(1-z_a/z_b)\ll 1$ and the main
contribution into the integral in Eq. (\ref{intTik1}) comes from
large values of $u$. Replacing the Bessel modified functions by
their asymptotic expansions for large values of the argument, for
the VEVs from Eq. (\ref{intTik1}) to the leading order one
receives
\begin{eqnarray}\label{TDD2plasymp}
\Delta \langle T_{D}^{D}\rangle &\sim & \frac{D\Gamma \left(
\frac{D+1}{2} \right) \zeta _R(D+1)}{(4\pi )^{(D+1)/2}(b-a)^{D+1}}
\frac{1-2^{-D-1}\left[ 1-\kappa (B_a)\kappa (B_b)\right] }{\kappa (B_a)
\kappa (B_b)} \\
\Delta \langle T_{0}^{0}\rangle &\sim & -\frac{1}{D}\Delta \langle
T_{D}^{D}\rangle -\frac{(\zeta -\zeta _c)}{2^{2D-1}\pi ^{D/2}
\Gamma (D/2)(b-a)^{D+1}} \times
\nonumber \\
&& \times \int _{0}^{\infty }\frac{t^Ddt}{\kappa (B_a)
\kappa (B_b)e^t-1}\left[ \kappa (B_a)\exp \left(
t\frac{z_a-z}{z_b-z_a}\right) +\kappa (B_b)\exp \left(
t\frac{z-z_b}{z_b-z_a}\right)\right] ,\label{Tii2plasymp}
\end{eqnarray}
where $\zeta _R(z)$ is the Riemann zeta function and the function
$\kappa (x)$ is defined by Eq. (\ref{kappa}). The leading terms
given by formulae (\ref{TDD2plasymp}) and (\ref{Tii2plasymp}) are
the same as for the corresponding quantities on Minkowski
spacetime background. In particular, the 'interference' part of
the ${}^{D}_{D}$ -- component does not depend on the curvature
coupling parameter. In the limit of large distances between the
plates, when $k_D(b-a)\gg 1$, introducing a new integration
variable $v=uz_b$ and expanding over $z_a/z_b$, to the leading
order from Eq. (\ref{intTik1}) one obtains
\begin{equation}\label{asympld}
\Delta \langle T_{i}^{k}\rangle \sim \delta _{i}^{k}\left(
\frac{z_a}{z_b}\right) ^{2\nu } g^{(i)}\left( \frac{z}{z_b}\right)
,
\end{equation}
where the form of the function $g^{(i)}(v)$ can be found from Eq.
(\ref{intTik1}). For $v\ll 1$ one has $g(v)\sim v^{D}$ and from
(\ref{asympld}) it follows that the 'interference' part of the
vacuum energy-momentum tensor is exponentially suppressed for
large interbrane distances.

\subsection{Interaction forces}

\begin{figure}[tbph]
\begin{center}
\begin{tabular}{cc}
\epsfig{figure=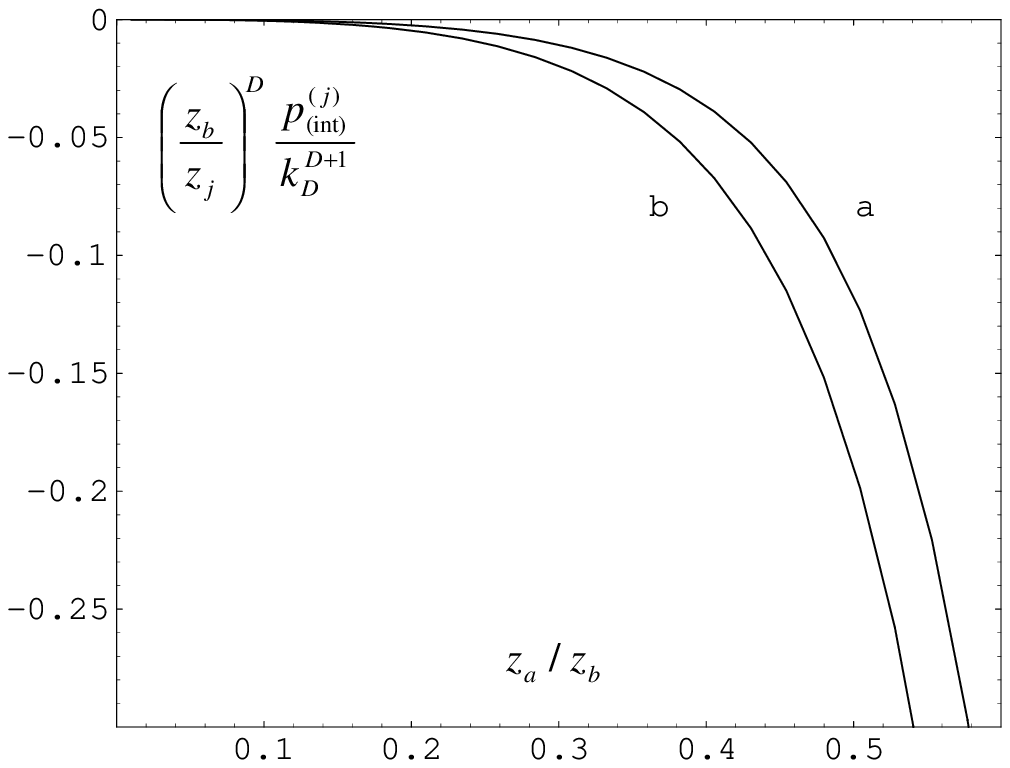,width=6.5cm,height=6cm}& \quad
\epsfig{figure=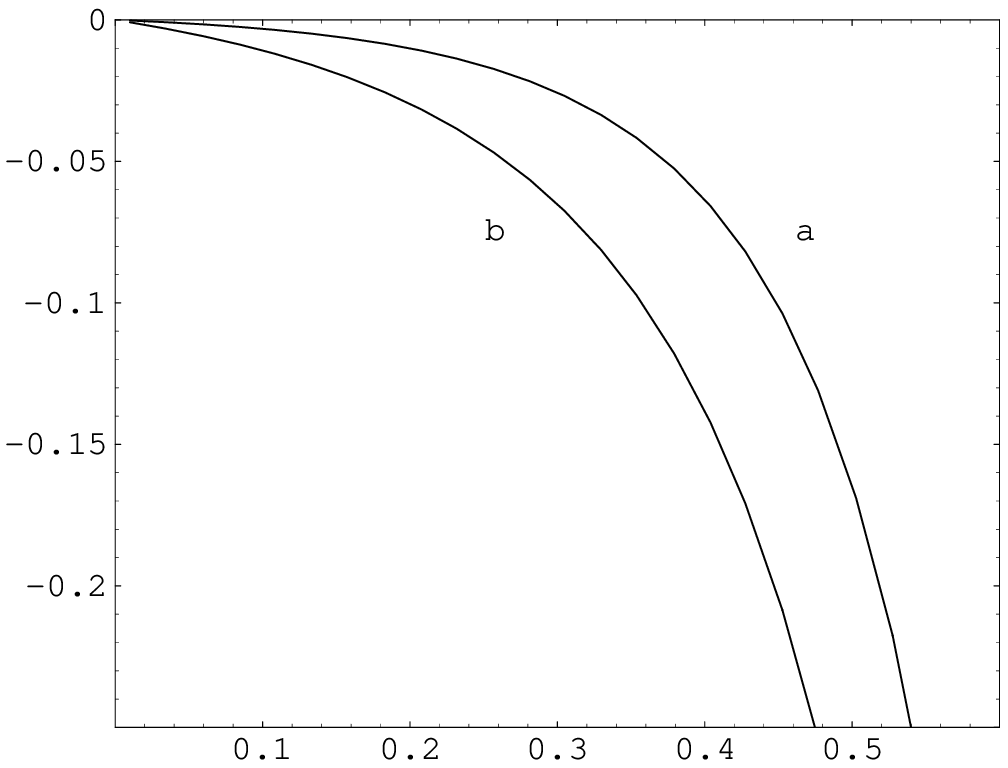,width=6.5cm,height=6cm}
\end{tabular}
\end{center}
\caption{The vacuum effective pressures
$(z_b/z_j)^{D}p^{(j)}_{{\mathrm{(int)}}}/k_D^{D+1}$, $j=a,b$, as
functions on $z_a/z_b$ for a massless minimally coupled scalar
field in $D=3$ determining the interaction forces between
Dirichlet (left panel) and Neumann (right panel) plates. Curves a
and b correspond to the pressures on the plates $z=z_j$, $j=a,b$
respectively.} \label{fig3dn}
\end{figure}

Now we turn to the interaction forces between the plates. The
vacuum force acting per unit surface of the plate at $z=z_j$ is
determined by the ${}^D_D$ -- component of the vacuum
energy-momentum tensor at this point. The corresponding effective
pressures can be presented as a sum of two terms:
\begin{equation}
p^{(j)}=p^{(j)}_1+p^{(j)}_{{\mathrm{(int)}}},\quad j=a,b. \label{pintdef}
\end{equation}
The first term on the right is the pressure for a single plate at
$z=z_j$ when the second plate is absent. This term is divergent
due to the surface divergences in the vacuum expectation values.
The second term on the right of Eq. (\ref{pintdef}),
\begin{equation}
p^{(j)}_{{\mathrm{(int)}}}=-\langle T_{D}^{D}\rangle ^{(j_1)}-
\Delta \langle T_{D}^{D}\rangle ,\quad z=z_j,\quad j,j_1=a,b,\quad j_1\neq j, \label{pintdef1}
\end{equation}
is the pressure induced by the presence of the second plate, and
can be termed as an interaction force. It is determined by the
last terms on the right of formulae (\ref{Tik1int}) and
(\ref{Tik1intb}) for the plate at $z=z_a$ and $z=z_b$
respectively. Substituting into these terms $z=z_j$ and using
relations
\begin{equation}
G_{\nu }^{(j)}(u,u)=-B_j,\quad G_{\nu }^{(j)'}(u,u)=A_j/u, \label{Guu}
\end{equation}
one has
\begin{eqnarray}
p^{(j)}_{{\mathrm{(int)}}}&=& \frac{k_D^{D+1}}{2^D\pi ^{D/2}
\Gamma \left( \frac{D}{2}\right)} \int _{0}^{\infty }dx\, x^{D-1}
\Omega _{j\nu }\left( x z_a/z_j,x z_b/z_j\right) \nonumber \\
&& \times \left\{ \left( x^2-\nu ^2 +2m^2/k_D^2\right) B_j^2 -
D(4\zeta -1) A_jB_j-A_j^2\right\} .\label{pintj}
\end{eqnarray}
Note that due to the asymmetry in the VEV of the energy-momentum
tensor, the interaction forces acting on the branes are not
symmetric under the interchange of the brane indices. By taking
into account that $K_{\nu }(u)I_{\nu }(v)- K_{\nu }(v)I_{\nu
}(u)>0$ and $K'_{\nu }(u)I'_{\nu }(v)- K'_{\nu }(v)I'_{\nu }(u)<0$
for $u<v$, it can be easily seen that the vacuum effective
pressures are negative for Dirichlet scalar and for a scalar with
$A_a=A_b=0$ and, hence, the corresponding interaction forces are
attractive for all values of the interplate distance. In figure
\ref{fig3dn} these forces are plotted as functions on $z_a/z_b$
for massless minimally coupled Dirichlet and Neumann scalars. As
it has been shown above, in the case of a conformally coupled
massless scalar field  for the first term on the right of Eq.
(\ref{pintdef}) one has $p_1^{(a)}=0$ and $p_1^{(b)}$ is
determined from Eq. (\ref{conf1pl}) with the replacement $a\to b$
(recall that we consider the region $z_a\leq z\leq z_b$). It can
be seen that the corresponding formulae for $p^{(j)}$ obtained
from Eqs. (\ref{pintdef}), (\ref{pintj}) coincide with those given
in Ref. \cite{Saha03a}. Note that in this case
$p^{(a)}/z_a^{D+1}=p^{(b)}/z_b^{D+1}$. Using the Wronskian for the
Bessel modified functions, it can be seen that
\begin{equation} \label{Omtrans}
\left[ B_j^2(x^2z_j^2+\nu ^2)-A_j^2\right] \Omega _{j\nu } (x z_a,x z_b)=
n_jz_j\frac{\partial }{\partial z_j}\ln \left| 1-
\frac{\bar I_{\nu }^{(a)}(xz_a)
\bar K_{\nu }^{(b)}(xz_b)}{\bar I_{\nu }^{(b)}(x
z_b)\bar K_{\nu }^{(a)}(xz_a)}\right| ,\quad j=a,b,
\end{equation}
where $n_a=1$, $n_b=-1$. This allows us to write the expressions
(\ref{pintj}) for the interaction forces per unit surface in
another equivalent form:
\begin{eqnarray}
p^{(j)}_{{\mathrm{(int)}}}&=& \frac{n_jk_D^{D+1}z_j^{D+ 1}}{2^D\pi
^{D/2}\Gamma \left( \frac{D}{2}\right)} \int _{0}^{\infty }dx\,
x^{D-1}\left[ 1+DB_j\frac{2
\zeta B_j+(1-4\zeta )\tilde A_j}{B_j^2(x^2z_j^2+\nu ^2)-A_j^2}\right] \nonumber \\
&& \times \, \frac{\partial }{\partial z_j}\ln \left| 1-
\frac{\bar I_{\nu }^{(a)}(xz_a)\bar K_{\nu }^{(b)}(x
z_b)}{\bar I_{\nu }^{(b)}(xz_b)\bar K_{\nu }^{(a)}(xz_a)}\right|  . \label{pint1}
\end{eqnarray}
Note that these forces in general are different for $j=a$ and $j=b$.
For Dirichlet scalar the second term in the square brackets is zero.

\begin{figure}[tbph]
\begin{center}
\begin{tabular}{cc}
\epsfig{figure=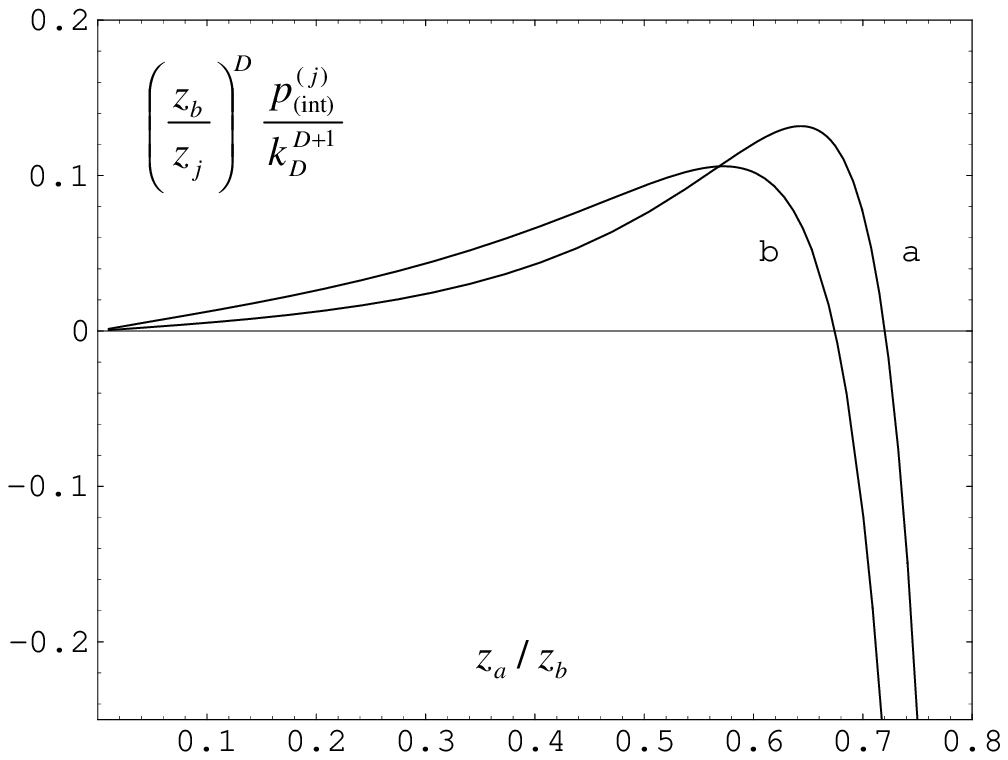,width=6.5cm,height=6cm}& \quad
\epsfig{figure=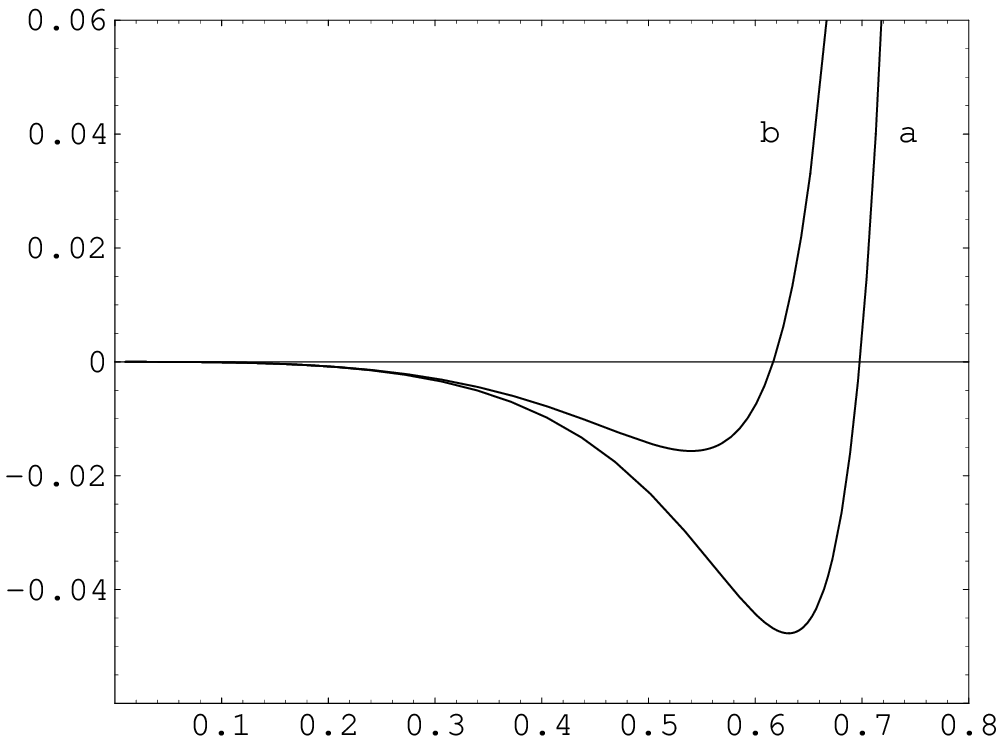,width=6.5cm,height=6cm}
\end{tabular}
\end{center}
\caption{The same as in figure \ref{fig3dn} for mixed boundary
conditions. Left panel corresponds to $\tilde A_a=0$, $B_a=1$,
$\tilde A_b=1$, $B_b=0.2$ and right panel corresponds to $\tilde
A_a=1$, $B_a=0$, $\tilde A_b=1$, $B_b=0.2$.} \label{fig4ro}
\end{figure}

\begin{figure}[tbph]
\begin{center}
\begin{tabular}{cc}
\epsfig{figure=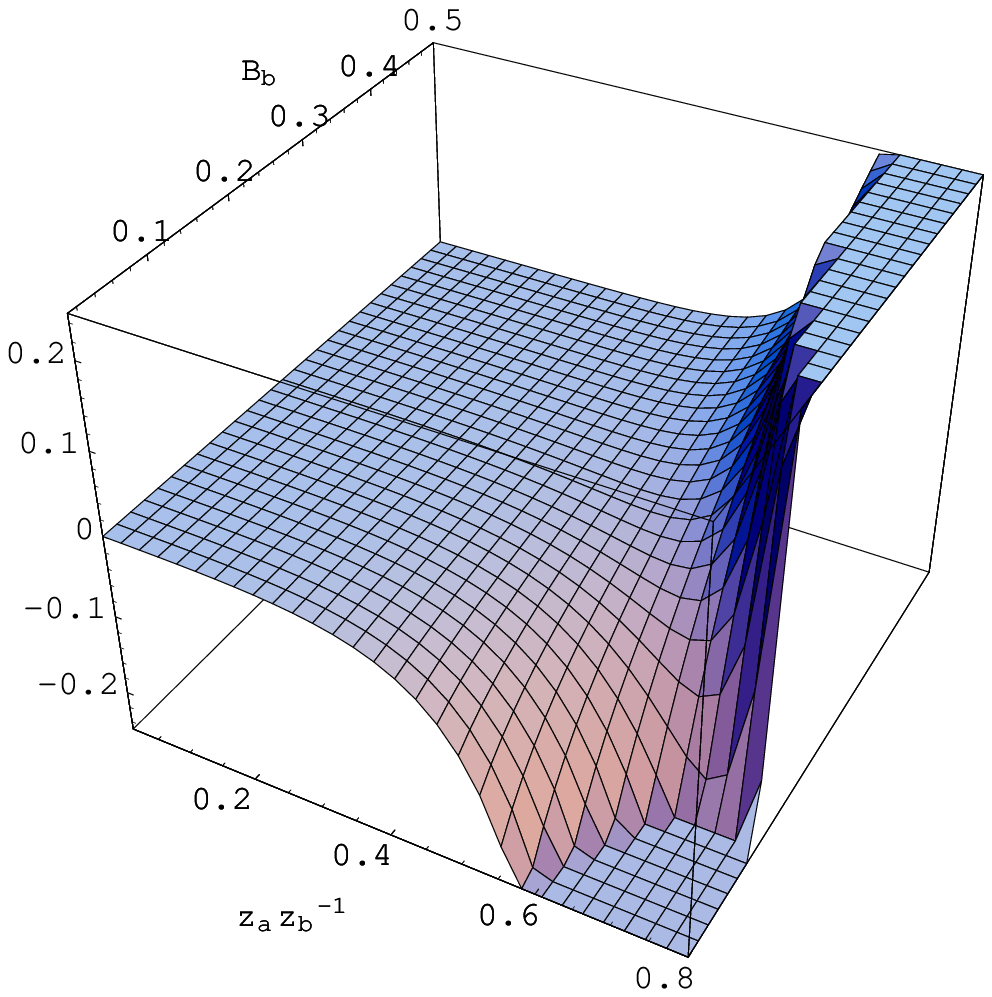,width=6.5cm,height=6.5cm}& \quad
\epsfig{figure=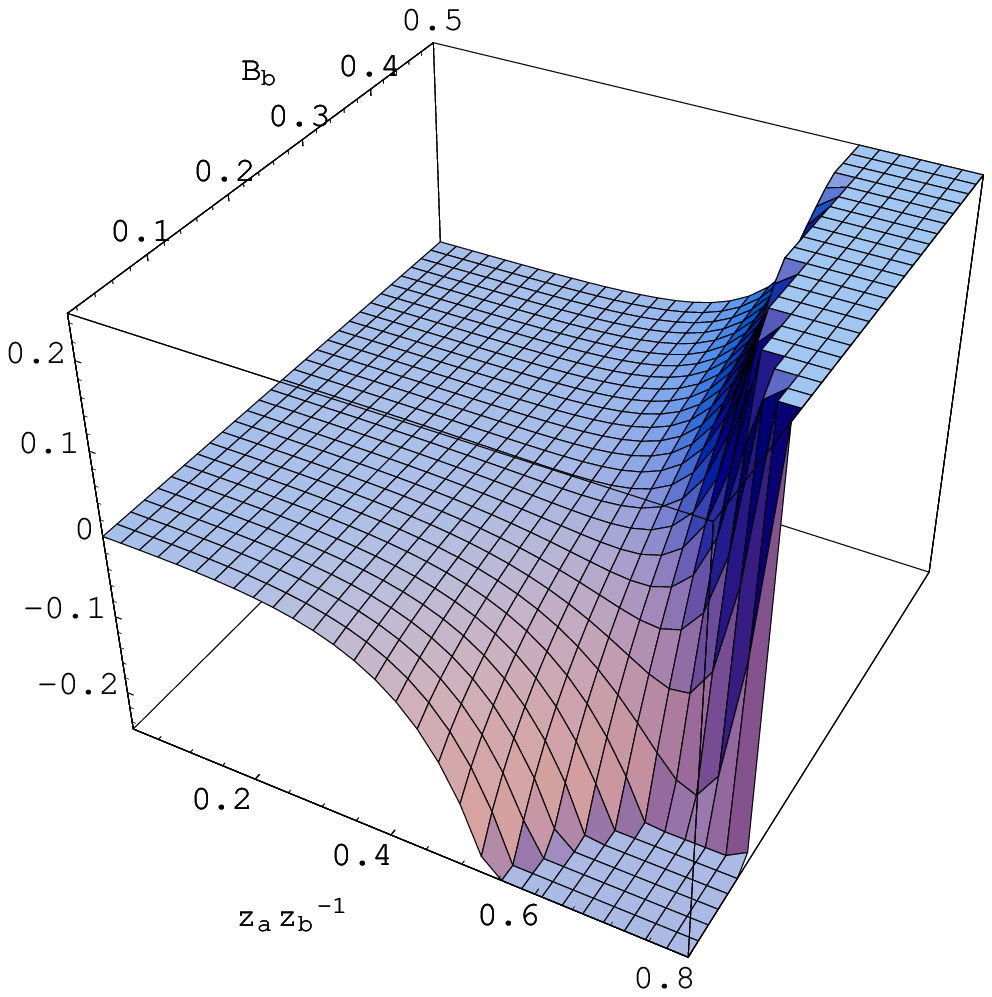,width=6.5cm,height=6.5cm}
\end{tabular}
\end{center}
\caption{The $D=3$ vacuum effective pressures $(z_b/z_j)^{D}p^{(j)}_{{\mathrm{int}}}/k_D^{D+1}$, $j=a,b$, for a massless minimally coupled scalar field as functions on $z_a/z_b$ and $B_b$. The values for the other Robin coefficients are $\tilde A_a=1$, $B_a=0$, and $\tilde A_b=1$. Left panel corresponds to $j=a$ and right panel corresponds to $j=b$.} \label{fig5ro}
\end{figure}

\begin{figure}[tbph]
\begin{center}
\begin{tabular}{cc}
\epsfig{figure=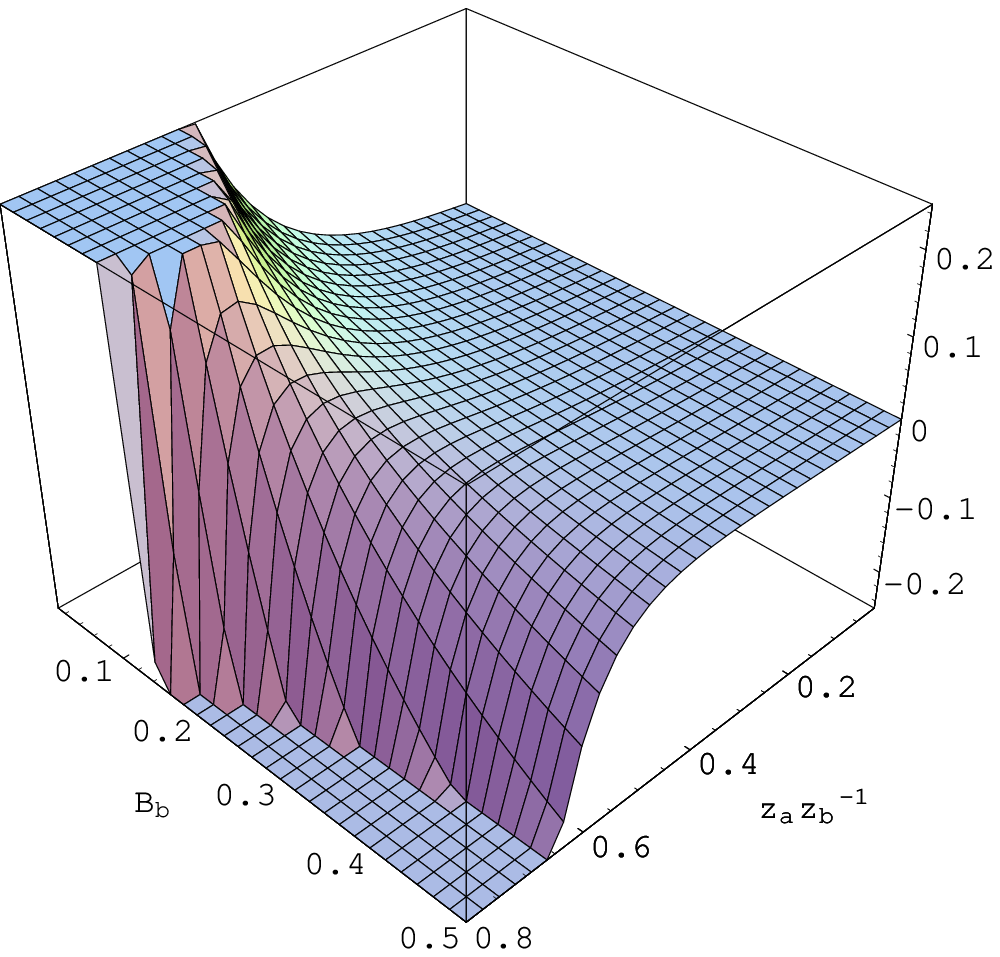,width=6.5cm,height=6cm}& \quad
\epsfig{figure=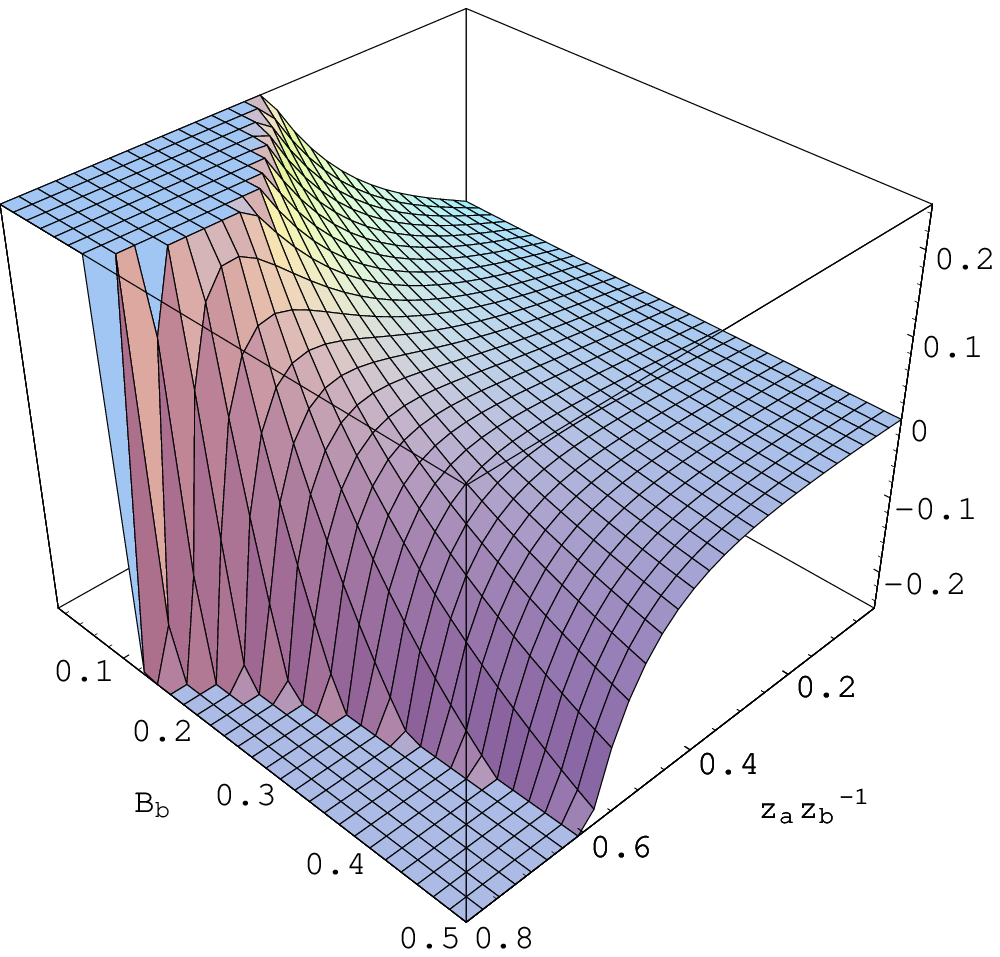,width=6.5cm,height=6cm}
\end{tabular}
\end{center}
\caption{The same as in figure \ref{fig5ro} for the values of
Robin coefficients $\tilde A_a=0$, $B_a=1$, and $\tilde A_b=1$.}
\label{fig6ro}
\end{figure}

Let us consider the limiting cases for the interaction forces
described by Eq. (\ref{pintj}). For small distances to compared
with the AdS curvature radius, $k_D(b-a)\ll 1$, the leading terms
are the same as for the plates in the Minkowski bulk:
\begin{equation} \label{forcelim1}
p^{(j)}_{{\mathrm{(int)}}}=D(1-2^{-D})\frac{\Gamma \left(
\frac{D+1}{2} \right) \zeta _R(D+1)}{(4\pi )^{(D+1)/2}(b-a)^{D+1}},
\end{equation}
in the case of Dirichlet boundary condition on one plate and
non-Dirichlet boundary condition on the another, and
\begin{equation} \label{forcelim2}
p^{(j)}_{{\mathrm{(int)}}}=-D\frac{\Gamma \left(
\frac{D+1}{2} \right) \zeta _R(D+1)}{(4\pi )^{(D+1)/2}(b-a)^{D+1}},
\end{equation}
for all other cases. Note that in the first case the interaction
forces are repulsive and are attractive for the second case. For
large distances between the plates, $k_D(b-a)\gg 1$ (this limit is
realized in the Randall-Sundrum model), by using the expressions
for the modified Bessel functions for small values of the
argument, one finds
\begin{eqnarray}
p^{(a)}_{{\mathrm{(int)}}}&\sim &  \left( \frac{z_a}{2z_b}\right) ^{D+
2\nu }\frac{4k_D^{D+1}\left[ (2m^2/k_D^2-\nu ^2)B_a^2-D(4\zeta -
1)A_aB_a-A_a^2\right]}{\pi ^{D/2}\Gamma (D/2) \Gamma ^2(\nu )(B_a\nu -A_a)^2} \nonumber \\
&& \times\int _{0}^{\infty }dx\, x^{2\nu +
D-1}\frac{\bar K_{\nu }^{(b)}(x)}{\bar I_{\nu }^{(b)}(x)} \, , \label{forcelarga}\\
p^{(b)}_{{\mathrm{(int)}}}&\sim &  \left( \frac{z_a}{2
z_b}\right) ^{2\nu } \frac{A_a+B_a\nu }{A_a-
B_a\nu }\frac{k_D^{D+1}}{2^{D-1}\pi ^{D/2}\Gamma (D/
2) \nu \Gamma ^2(\nu )} \nonumber \\
&& \times \int _{0}^{\infty }dx\, \frac{x^{2\nu + D-1}}{\bar
I_{\nu }^{(b)2}(x)} \left[ (x^2-\nu ^2+ 2m^2/k_D^2)B_b^2-D(4\zeta
-1) A_bB_b-A_b^2\right] .\label{forcelargb}
\end{eqnarray}
In dependence of the values for the coefficients in the boundary
conditions, these effective pressures can be either positive or
negative, leading to repulsive or attractive forces at large
distances. To illustrate these possibilities, in figure
\ref{fig4ro} we have plotted the effective pressures as functions
on $z_a/z_b$ for Robin boundary conditions. In the case of the
left panel, the vacuum interaction forces are attractive for small
distances and repulsive for large distances. For the right panel
we have an opposite situation, the interaction forces are
attractive for large distances and repulsive for small distances.
For the latter case the vacuum interaction forces provide a
possibility for a stabilization of the interplate distance. The
dependence of the vacuum effective pressures on the ratio
$z_a/z_b$ for various values of the Robin coefficient $B_b$ is
given in figures \ref{fig5ro} and \ref{fig6ro}.

\section{Application to the Randall-Sundrum braneworld} \label{sec:RanSum}

In this section we will consider the application of the results
obtained in the previous sections to the Randall-Sundrum
braneworld model \cite{Rand99a} based on the AdS geometry with one
extra dimension. The fifth dimension $y$ is compactified on an
orbifold, $S^1/Z_2$ of length $L$, with $-L\leq y\leq L$. The
orbifold fixed points at $y=0$ and $y=L$ are the locations of two
3-branes. Below we will allow these submanifolds to have an
arbitrary dimension $D$. The metric in the Randall-Sundrum model
has the form (\ref{metric}) with
\begin{equation}\label{sigmaRS}
  \sigma (y)=k_D|y|.
\end{equation}
For the corresponding Ricci scalar one has
\begin{equation}\label{RicciRS}
  R=4Dk_D[\delta (y)-\delta (y-L)]-D(D+1)k_D^2.
\end{equation}
Let us consider a bulk scalar $\varphi $ with the curvature
coupling parameter $\zeta $ and with the action functional
\begin{equation}\label{actionsc}
S=\frac{1}{2}\int d^Dxdy\sqrt{|g|}\left\{ g^{ik}\partial _i
\varphi \partial _k \varphi -\left[ m^2 +c_1\delta (y)+c_2\delta
(y-L)+\zeta R \right] \varphi ^2 \right\} ,
\end{equation}
where $m$ is the bulk mass, $c_1$ and $c_2$ are the brane mass
terms on the branes $y=0$ and $y=L$ respectively. The field
equation obtained from (\ref{actionsc}) is
\begin{equation}\label{fieldeqRS}
  \nabla _i\nabla ^i \varphi +\left[ m^2 +c_1\delta (y)+c_2\delta
(y-L)+\zeta R \right] \varphi =0 .
\end{equation}
The corresponding eigenfunctions can be written in form
(\ref{eigfunc1}) and (\ref{branefunc1}), where $f_n(y)$ is a
solution to the equation
\begin{eqnarray}\label{egforfnRS}
  -e^{D\sigma }\frac{d}{dy}\left( e^{-D\sigma
  }\frac{df_n}{dy}\right) & + & \left[ m^2+(c_1+4D\zeta k_D)\delta (y)+
  (c_2-4D\zeta k_D)\delta (y-L) -\right. \nonumber \\
  && \left. -D(D+1)\zeta k_D^2\right] f_n(y)=
  m_n^2e^{2\sigma }f_n(y) .
\end{eqnarray}
To obtain the boundary conditions for the function $f_n(y)$ we
integrate (\ref{egforfnRS}) first about $y=0$ and then about
$y=L$. Assuming that the function $f_n(y)$ is continuous at these
points, we receive
\begin{equation} \label{boundRS1}
\lim _{\epsilon \to 0}\frac{df_{n}(y)}{dy}\left| _{y=-y_j+(-1)^j\epsilon
}^{y=y_j-(-1)^j\epsilon }\right. =\left[ c_j-(-1)^j 4D\zeta k_D\right]
f_n(y_j),\quad j=1,2,\quad y_1=0,\quad y_2=L \, ,
\end{equation}
with $\epsilon >0$. First consider the case of the untwisted
scalar field for which  $f_n(y)=f_n(-y)$, and the solution to Eq.
(\ref{egforfnRS}) for $y\neq 0,L$ is given by expression
(\ref{fny}) with $z=e^{k_D|y|}$. The boundary conditions which
follow from relations (\ref{boundRS1}) are in form
(\ref{boundcond}) with (see also Ref. \cite{Flac01b} for the case
$c_1=c_2=0$)
\begin{equation}\label{AtildeRS}
  \frac{\tilde A_a}{\tilde B_a} =  -\frac{1}{2}(c_1+4D\zeta k_D),\quad
\frac{\tilde A_b}{\tilde B_b} = -\frac{1}{2}(-c_2+4D\zeta k_D),
\end{equation}
and respectively
\begin{equation}\label{ARS}
\frac{A_a}{B_a} = \frac{1}{2}\left[ D(1-4\zeta )-c_1/k_D\right] ,\quad
\frac{A_b}{B_b} = \frac{1}{2}\left[ D(1-4\zeta )+c_2/k_D\right] .
\end{equation}
For the twisted scalar $f_n(-y)=-f_n(y)$ and from the $Z_2$
identification on $S^1$ one has $f_n(0)=f_n(L)=0$ and, hence,
$\tilde B_a=\tilde B_b=0$, which correspond to Dirichlet boundary
conditions on both branes. The normalization condition for the
eigenfunctions $f_n(y)$ now has the form
\begin{equation}\label{normcondRS}
  \int_{-L}^{L}dye^{(2-D)\sigma (y)}f_n(y)f_{n'}(y)=\delta _{nn'}.
\end{equation}
As a result the normalization coefficient $c_n$ differs from
(\ref{cnu}) by additional factor 1/2. Hence, the Wightman function
in the Randall-Sundrum braneworld is given by formulae (\ref{W15})
or equivalently by (\ref{W17}) with an additional factor 1/2 and
with Robin coefficients given by Eq. (\ref{AtildeRS}). Similarly,
the VEV of the energy-momentum tensor is defined by formulae
(\ref{Tik1int}) or (\ref{Tik1intb}) by an additional factor 1/2.
Note that global quantities, such as the total Casimir energy, are
the same. Motivated by the possibility for the stabilization of
extra dimensions by quantum effects, the one-loop Casimir energy
of various braneworld compactifications has been investigated by
several authors for both scalar and fermion fields (see references
cited in Introduction). It has been shown that for the
simultaneous solution of the stabilization and hierarchy problems
a fine tuning of the model parameters is essential. From the point
of view of backreaction of quantum effects the investigation of
local densities is also of considerable interest. For a
conformally coupled massless scalar field the corresponding
results can be obtained from the results in the Minkowski
spacetime by using the standard formula for conformally related
problems. For the case of plane-symmetric line element
(\ref{metric}) with an arbitrary function $\sigma (y)$ this has
been done in Ref. \cite{Saha03a} by using the results from
\cite{Rome02}. Recently the energy-momentum tensor in the
Randall-Sundrum braneworld for a bulk scalar with zero brane mass
terms $c_1$ and $c_2$ is considered in Ref. \cite{Knap03}. This
paper appeared when our calculations were in progress. Note that
in \cite{Knap03} only a general formula is given for the
unrenormalized VEV in terms of the differential operator acting on
the Green function. In our approach the application of the
generalized Abel-Plana formula allowed us to extract manifestly
the part due to the AdS bulk without boundaries and for the points
away from the boundaries the renormalization procedure is the same
as for the boundary-free parts. The latter is well-investigated in
literature. In addition, the boundary induced parts are presented
in terms of exponentially convergent integrals convenient for
numerical calculations.

\section{Conclusion} \label{sec:Conc}

The natural appearance of AdS in a variety of situations has
stimulated considerable interest in the behavior of quantum fields
propagating in this background. In the present paper we have
investigated the Wightman function and the vacuum expectation
value of the energy-momentum tensor for a scalar field with an
arbitrary curvature coupling parameter satisfying Robin boundary
conditions on two parallel plates in AdS spacetime. The
application of the generalized Abel-Plana formula to the mode sum
over zeros of the corresponding combinations of the cylinder
functions allowed us to extract the boundary-free AdS part and to
present the boundary induced parts in terms of exponentially
convergent integrals. In the region between two plates the
Wightman function is presented in two equivalent forms given by
Eqs. (\ref{W15}) and (\ref{W17}). The first terms on the right of
these formulae are the Wightman function for AdS bulk without
boundaries. The second ones are induced by a single plate and the
third terms are due to the presence of the second plate. The
expectation values for the energy-momentum tensor are obtained by
applying on the Wightman function a certain second order
differential operator and taking the coincidence limit. For the
case of a single plate geometry this leads to formula
(\ref{Tik1plnew}) for the region $z>z_a$ and to formula
(\ref{Tik1plnewleft}) for the region $z<z_a$. As we could expect
from the problem symmetry the part of this tensor corresponding to
the components on the hyperplane parallel to the plate is
proportional to the corresponding metric tensor. On the boundary
the vacuum energy-momentum tensor diverges, except the case of a
conformally coupled massless scalar. The leading terms of the
corresponding asymptotic expansion near the boundary are given by
expressions (\ref{Tik1near}) and (\ref{TDD1near}). These terms are
the same as for a plate in the Minkowski bulk. They do not depend
on the Robin coefficient, have different signs for Dirichlet and
Neumann scalars and vanish for a conformally coupled scalar. The
coefficients for the subleading asymptotic terms will depend on
the AdS curvature radius, Robin coefficient and on the mass of the
field. For large proper distances from the plate to compared with
the AdS curvature radius, $k_D|y-a|\gg 1$, the boundary induced
energy-momentum tensor vanishes as $\exp [2\nu k_D(a-y)]$ in the
region $y>a$ and as $\exp [k_D(2\nu +D)(y-a) ]$ in the region
$y<a$. The same behavior takes place for a fixed $y-a$ and large
values of the parameter $k_D$. In the large mass limit, $m\gg
k_D$, the boundary parts are exponentially suppressed. Note that
here we consider the bulk energy-momentum tensor. For a scalar
field on manifolds with boundaries in addition to this part, the
energy-momentum tensor contains a contribution located on on the
boundary (for the expression of the surface energy-momentum tensor
in the case of arbitrary bulk and boundary geometries see Ref.
\cite{Saha03}). As it has been discussed in Refs.
\cite{Kenn80,Rome02,Sahsph,Rome01,Full03,Saha03}, the surface part
of the energy-momentum tensor is essential in considerations of
the relation between local and global characteristics in the
Casimir effect. The vacuum expectation value of the surface
energy-momentum tensor for the geometry of two parallel branes in
AdS bulk is evaluated in Ref. \cite{Saha04b}. It is shown that for
large distances between the branes the induced surface densities
give rise to an exponentially suppressed cosmological constant on
the brane. In particular, in the Randall-Sundrum model the
cosmological constant generated on the visible brane is of the
right order of magnitude with the value suggested by the
cosmological observations.

For the case of two-plates geometry the vacuum expectation value
of the bulk energy-momentum tensor is presented as a sum of purely
AdS, single plates, and 'interference' parts. The latter is given
by Eq. (\ref{intTik1}) and is finite for all values $z_a\le z\le
z_b$. In the limit $z_a\to z_b$ the standard result for two
parallel plates in the Minkowski bulk is obtained. For two-plates
case the vacuum forces acting on boundaries contain two terms. The
first ones are the forces acting on a single boundary when the
second boundary is absent. Due to the well-known  surface
divergencies in the expectation values of the energy-momentum
tensor these forces are infinite and need an additional
regularization. The another terms in the vacuum forces are finite
and are induced by the presence of the second boundary and
correspond to the interaction forces between the plates. These
forces are given by formula (\ref{pintj}) with $j=a,b$ for the
plate at $z=z_a$ and $z=z_b$ respectively. For Dirichlet scalar
they are always attractive. In the case of mixed boundary
conditions the interaction forces can be either repulsive or
attractive. Moreover, there is a region in the space of Robin
parameters in which the interaction forces are repulsive for small
distances and are attractive for large distances. This provides a
possibility to stabilize interplate distance by using the vacuum
forces. For large distances between the plates, the vacuum
interaction forces per unit surface are exponentially suppressed
by the factor $\exp [2\nu k_D(a-b)]$ for the plate at $y=a$ and by
the factor $\exp [k_D(2\nu +D)(a-b) ]$ for the plate at $y=b$. In
Section \ref{sec:RanSum} we give an application of our results for
two plates case to the Randall-Sundrum braneworld with arbitrary
mass terms on the branes. For the untwisted scalar the Robin
coefficients are expressed through these mass terms and the
curvature coupling parameter. For the twisted scalar Dirichlet
boundary conditions are obtained on both branes. Note that in this
paper we have considered boundary induced vacuum densities which
are finite away from the boundaries. As it has been mentioned in
Ref. \cite{Grah03}, the same results will be obtained in the model
where instead of externally imposed boundary condition the
fluctuating field is coupled to a smooth background potential that
implements the boundary condition in a certain limit
\cite{Grah02}.

\section*{Acknowledgments}

I am grateful to Michael Bordag and Dmitri Vassilevich for
valuable discussions and suggestions. The work was supported by
the DAAD grant. I acknowledge the hospitality of the Institute for
Theoretical Physics, University of Leipzig.

\end{document}